%% file: Main.tex
\documentclass[acmsmall,screen]{acmart}

\AtBeginDocument{%
  }

\copyrightyear{2025}
\acmYear{2025}

\acmJournal{TOSEM}
\acmVolume{1}
\acmNumber{1}
\acmArticle{1}
\acmMonth{1}

\usepackage{natbib}

\usepackage{amsmath}

\usepackage{amssymb}
\usepackage{amsfonts}

\usepackage{adjustbox}
\usepackage{graphicx}
\usepackage{textcomp}
\usepackage{xcolor}

\usepackage{xspace}
\usepackage{multirow}
\usepackage{framed}
\usepackage{tcolorbox}
\usepackage{diagbox}
\usepackage{pifont}
\usepackage{subcaption}
\usepackage{orcidlink}

\usepackage{colortbl}
\usepackage{url} 
\usepackage{listings}

\usepackage{algorithmic}
\usepackage[linesnumbered,ruled,vlined]{algorithm2e}

\usepackage{enumitem}

\usepackage{float}
\newfloat{codelist}{tbhp}{lop}
\floatname{codelist}{Listing}

\begin{document}

\input{macros}

\title{\ourtool: Fuzzing Compilers via LLM-Synthesized Mutators Learned from Bug Reports}

\author{Bo Wang}
\orcid{0000-0001-7944-9182}
\affiliation{
  \institution{Beijing Jiaotong University}
  \city{Beijing}
  \country{China}
}
\email{wangbo\_cs@bjtu.edu.cn}

\author{Pengyang Wang}
\orcid{0009-0005-0387-826X}
\affiliation{
  \institution{Beijing Jiaotong University}
  \city{Beijing}
  \country{China}
}
\email{25125334@bjtu.edu.cn}

\author{Chong Chen}
\orcid{0009-0004-0721-2943}
\affiliation{
  \institution{Beijing Jiaotong University}
  \city{Beijing}
  \country{China}
}
\email{22120350@bjtu.edu.cn}

\author{Ming Deng}
\orcid{0009-0007-7193-0087}
\affiliation{
  \institution{Beijing Jiaotong University}
  \city{Beijing}
  \country{China}
}
\email{24120317@bjtu.edu.cn}

\author{Jieke Shi}
\orcid{0000-0002-0799-5018}
\affiliation{
  \institution{Singapore Management University}
  \country{Singapore}
}
\email{jiekeshi@smu.edu.sg}

\author{Qi Sun}
\orcid{0009-0006-5116-4752}
\affiliation{
  \institution{Beijing Jiaotong University}
  \city{Beijing}
  \country{China}
}
\email{21281049@bjtu.edu.cn}

\author{Chengran Yang}
\orcid{0000-0001-6100-8127}
\affiliation{
  \institution{Singapore Management University}
  \country{Singapore}
}
\email{cryang@smu.edu.sg}

\author{Zhou Yang}
\orcid{0000-0001-5938-1918}
\affiliation{
  \institution{University of Alberta}
  \city{Alberta}
  \country{Canada}
}
\email{zy25@ualberta.ca}

\author{Youfang Lin}
\orcid{0000-0002-5143-3645}
\affiliation{
  \institution{Beijing Jiaotong University}
  \city{Beijing}
  \country{China}
}
\email{yflin@bjtu.edu.cn}

\author{Junjie Chen}
\orcid{0000-0003-3056-9962}
\affiliation{
  \institution{Tianjin University}
  \city{Tianjin}
  \country{China}
}
\email{junjiechen@tju.edu.cn}

\author{Jun Sun}
\orcid{0000-0002-3545-1392}
\affiliation{
  \institution{Singapore Management University}
  \country{Singapore}
}
\email{junsun@smu.edu.sg}

\author{David Lo}
\orcid{0000-0002-4367-7201}
\affiliation{
  \institution{Singapore Management University}
  \country{Singapore}
}
\email{davidlo@smu.edu.sg}

\renewcommand{\shortauthors}{Wang et al.}

\begin{abstract}
Mutation-based fuzzing has proven effective in uncovering compiler bugs. However, designing high-quality mutation operators (i.e., mutators) for modern programming languages that feature complex language constructs (e.g., templates and macros) remains challenging. Existing methods rely heavily on manual effort, either to craft a limited set of simple mutators or to inspect and correct automatically generated ones. This reliance limits the expressiveness of mutators, hampers scalable mutator synthesis, and consequently restricts cross-language generalizability as mutators are often tailored to specific languages.

In this work, we present \ourtool, \bocomment{a cross-language framework for automated mutator synthesis that leverages} LLM agents and compiler-specific insights extracted from bug reports on open-source compiler repositories.
It is a \textit{multi-agent framework} comprising three specialized LLM agents: (1)~\textit{a mutator invention agent}, which identifies mutation targets and generates mutator names, signatures, and descriptions based on compiler-related insights (e.g., language features explicitly mentioned in bug reports); (2)~\textit{a mutator implementation synthesis agent}, which is enhanced via fine-tuning to generate initial mutator implementations; and (3)~\textit{a mutator refinement agent}, which verifies and corrects the synthesized mutators using feedback from unit tests.
\bocomment{To guide the LLM toward more accurate and up-to-date use of AST modification APIs}, we fine-tune the LLM of the latter two agents using a small set of human-written mutators.
Additionally, we propose an adaptive seed enhancement approach to generate more diverse seed programs.

\bocomment{In our main evaluation, \ourtool} analyzes 1000 bug reports (500 each for Rust and C++), and synthesizes 319 and 403 mutators for them. 
\bocomment{The average end-to-end API cost per final valid mutator is approximately \$0.08 using GPT-4o, including mutator invention, implementation synthesis, and refinement attempts.
To further study the impact of the bug-report budget, we analyze an additional 200 reports per language and update the bug-hunting campaign accordingly. Using the mutators synthesized from 700 reports per language, our customized fuzzer reported 100 bugs across four production compilers, 61 of which have been confirmed as previously unknown by developers, and 35 of which have already been fixed.}
Experimental results also show that \ourtool discovers a significant number of unique crashes, complementing existing fuzzers, and achieves the highest coverage on all Rust and C++ compilers.

\end{abstract}

\ccsdesc[500]{Software and its engineering~Software testing and debugging}
\ccsdesc[500]{Software and its engineering~Compilers}

\keywords{Mutation-Based Compiler Fuzzing, LLM Agents, Compiler Bugs, Rust Compiler, C++ Compiler}

\maketitle

\input{Intro}
\input{Motivating}
\input{Approach}
\input{Eval}

\input{Discussion}
\input{Related}

\input{Conclusion}

\bibliographystyle{ACM-Reference-Format}
\bibliography{ref}

\end{document}

%% file: macros.tex
\newcommand{\code}[1]{\texttt{#1}}
\newcommand{\todo}[1]{{\color{red} \bf \{TODO: {#1}\}}}

\newcommand{\wb}[1]{{\color{brown}[Bo: #1]}}

\def\modify#1#2#3{{\small{\sf{#1}}} {\color{red}{#2}}
{{\color{red}\mbox{$\Rightarrow$}}} {\color{blue}{#3}}}

\newcommand{\mycomment}[2]{{\color{magenta}{\sf{#1}}} {\color{blue}{#2}}}
\renewcommand{\mycomment}[2]{{#2}}

\newcommand{\bocomment}[1]{\mycomment{}{#1}}
\newcommand{\bomodify}[2]{\modify{}{#1}{#2}}
\newcommand{\bomodifyno}[2]{{#1}}

\newcommand{\ourtoolraw}{Mut4All}
\newcommand{\ourtool}{\ourtoolraw{}\xspace}

\newcommand{\listref}[1]{List~\ref{lst:#1}}
\newcommand{\tabref}[1]{Table~\ref{tab:#1}}
\newcommand{\tablabel}[1]{\label{tab:#1}}
\newcommand{\figref}[1]{Figure~\ref{fig:#1}}
\newcommand{\figlabel}[1]{\label{fig:#1}}

\newcommand{\smalltitle}[1]{{\smallskip \noindent \bf  {#1}.\ }}
\newcommand{\smalltitlecolon}[1]{{\smallskip \noindent \bf  {#1}:\ }}

\newcommand{\finding}[1]{ \begin{tcolorbox}
\textbf{Finding \refstepcounter{num}\thenum}: #1
\end{tcolorbox}}

\newcounter{num}

\lstdefinelanguage{C++}{
  morekeywords={asm, auto, bool, break, case, catch, char, class, const, const_cast,
    continue, default, delete, do, double, dynamic_cast, else, enum, explicit, export,
    extern, false, float, for, friend, goto, if, inline, int, long, mutable, namespace,
    new, operator, private, protected, public, register, reinterpret_cast, return, short,
    signed, sizeof, static, static_cast, struct, switch, template, this, throw, true,
    try, typedef, typeid, typename, union, unsigned, using, virtual, void, volatile,
    wchar_t, while},
  sensitive=true,
  morecomment=[l]{//},
  morecomment=[s]{/*}{*/},
  morestring=[b]",
  morestring=[b]',
}

\lstdefinestyle{mystyleCPP}{
  language=C++,
  basicstyle=\ttfamily\scriptsize,
  keywordstyle=\color{blue}\bfseries,
  commentstyle=\color{green!60!black},
  stringstyle=\color{red!70!black},
  numbers=left,
  numberstyle=\tiny\color{gray},
  stepnumber=1,
  numbersep=5pt,
  tabsize=2,
  showspaces=false,
  showstringspaces=false,
  breaklines=true,
}

\lstdefinelanguage{Rust}{
  morekeywords=[1]{fn, let, mut, if, else, match, loop, while, for, in, break, continue, return, impl, struct, enum, trait, mod, pub, use, crate, super, Self, self, ref, as, const, static, where, move, async, await, dyn, unsafe},
  sensitive=true,
  morecomment=[l]{//},
  morecomment=[s]{/*}{*/},
  morestring=[b]{"},
}

\lstdefinestyle{mystyleRust}{
  language=Rust,
  basicstyle=\ttfamily\scriptsize,
  keywordstyle=\color{blue}\bfseries,
  commentstyle=\color{green!60!black},
  stringstyle=\color{red!70!black},
  numbers=left,
  numberstyle=\tiny\color{gray},
  stepnumber=1,
  numbersep=5pt,
  tabsize=2,
  showspaces=false,
  showstringspaces=false,
  breaklines=true,
}

\lstdefinestyle{mystyleCpp}{
  language=C++,
  basicstyle=\ttfamily\scriptsize,
  keywordstyle=\color{blue}\bfseries,
  commentstyle=\color{green!60!black},
  stringstyle=\color{red!70!black},
  numbers=left,
  numberstyle=\tiny\color{gray},
  stepnumber=1,
  numbersep=5pt,
  tabsize=2,
  showspaces=false,
  showstringspaces=false,
  breaklines=true,
}

%% file: Intro.tex
\section{Introduction}

Compiler fuzzing~\cite{chen2020survey,marcozzi2019compiler,chen2016empirical} has proven effective in improving the reliability of production compilers such as GCC and rustc, uncovering numerous bugs over the past decades. 
\bocomment{Existing compiler fuzzers can be broadly categorized into generation-based and mutation-based approaches, although this is only a coarse-grained classification. Generation-based fuzzers construct test programs from scratch, ranging from grammar-based generators such as CSmith~\cite{yang2011finding} and YARPGen~\cite{livinskii2020random,livinskii2023fuzzing} to more semantics-aware generators that encode additional constraints such as typing rules~\cite{dewey2015fuzzing}. Such techniques can generate valid programs when sufficient semantic knowledge is encoded. 
However, for modern languages with complex static semantics and compiler-specific features, building and maintaining these semantic models, templates, or rule systems often requires substantial manual effort. 
This limits their scalability and adaptability when targeting new languages or advanced language constructs.}

In contrast, mutation-based fuzzers such as GrayC~\cite{even2023grayc}, MetaMut~\cite{ou2024mutators}, and TyMut~\cite{wang2025fuzzing}, which generate new test programs by applying mutation operators (i.e., mutators) to existing real-world programs (i.e., seed programs), are often preferred for compiler fuzzing, as they leverage existing code to avoid the need for designing complex templates and preserve both the syntactic and semantic validity of test programs. That being said, their effectiveness heavily depends on the quality and expressiveness of the mutators. Moreover, existing mutation-based fuzzers still face several major challenges in mutator design and implementation:

\textit{C1: Limited Expressiveness.} Most existing mutation-based fuzzers define mutators as compositions of simple operators, such as arithmetic adjustments (e.g., replacing ``\texttt{+}" with ``\texttt{-}") or control-flow rewrites (e.g., converting \texttt{switch} statements into \texttt{if-else} chains), as exemplified by GrayC~\cite{even2023grayc} and, to a more advanced extent, MetaMut~\cite{ou2024mutators}. While effective at exposing bugs in basic language constructs, these mutators inherently lack expressiveness—the ability to construct program variants that incorporate advanced language features and trigger deeper compiler logic, such as trait objects in Rust or templates in C++, both of which have received limited attention in prior studies~\cite{yang2024rust,even2023grayc,ou2024mutators,wang2024rustlantis,liu2025empirical}. This limitation often results in missed opportunities to uncover subtle bugs that arise from intricate interactions between language features, which are prevalent in realistic programs~\cite{wang2025comprehensive}.

\textit{C2: Significant Manual Effort.}
Existing mutation-based compiler fuzzers are mostly semi-automatic and still require active involvement from compiler experts. Tools like GrayC~\cite{even2023grayc} and DIE~\cite{park2020fuzzing} rely on manually crafted mutators, incurring high engineering costs—the former implements 13 mutators using over 8,000 lines of C++, while the latter requires more than 16,000 lines for its mutation engine. 
More recent approaches leverage Large Language Models (LLMs), which have demonstrated strong capabilities across programming tasks~\cite{hou2024large,wang2024software,chen2025deep}, for mutator generation.
MetaMut~\cite{ou2024mutators} guides the LLM to assemble mutators from around 100 predefined AST-level operations, while Clozemaster~\cite{gao2025clozemaster} fine-tunes LLMs to complete masked syntax elements.
While this significantly reduces manual implementation, it still involves non-trivial effort in prompt engineering, as well as validating and refining the generated code. 
For instance, MetaMut reports over two weeks of manual effort to repair and refine just 68 mutators~\cite{ou2024mutators}. 
Although promising, these methods still have room for improvement in scalability and automation, especially when scaling to support a large number of mutators for diverse compiler features.

\textit{C3: Limited Cross-Language Generalizability.}
Ideally, a compiler fuzzer should work across multiple programming languages and compilers for broader applicability~\cite{xia2024fuzz4all}. That said, due to substantial differences in language design and compiler infrastructures, mutators' implementation is often tightly coupled to a specific compiler. Even when designed to be generic, mutators that work well for one compiler may not be effective—or even functional—for another, making cross-language reuse difficult. Most existing fuzzers are designed for a single language, with mutators tailored to specific compiler infrastructures (e.g., GrayC~\cite{even2023grayc} and Clozemaster~\cite{gao2025clozemaster} only work for C and Rust, respectively). Extending them to new targets is inherently challenging, often requiring thousands of lines of code to accommodate differences in syntax, type systems, and compiler internals. While the rise of LLMs trained on multilingual code offers promise for cross-language transfer~\cite{xia2024fuzz4all}, current mutation-based fuzzers (even MetaMut and Clozemaster) have yet to fully exploit this capability. Achieving high-quality mutator transfer and reuse across languages remains an open challenge.

\smalltitle{Our work}
We introduce \ourtool, \bocomment{a cross-language framework for automated mutation-operator synthesis} to support mutation-based compiler fuzzing. The core idea behind \ourtool is to automate the entire mutator lifecycle, from discovery and design to implementation, by coordinating multiple specialized LLM agents that extract compiler-specific insights from historical bug reports in open-source repositories (e.g., GitHub). These real-world bug reports help identify bug-prone language features and provide concrete test programs that reveal compiler failures, allowing LLMs to focus on patterns that have previously caused actual bugs and produce more targeted and effective mutations. At a high level, \ourtool (1)~analyzes real-world bug reports to identify language features prone to compiler bugs and formulates corresponding mutator specifications, (2)~synthesizes concrete mutator implementations based on these specifications, and (3)~validates and refines the generated mutators to ensure their correctness and effectiveness.
In addition, we propose an adaptive seed enhancement technique aimed at improving the diversity of the original seed corpus, which includes both historical bug reports and official test cases.

Concretely, given a target compiler, the workflow begins by automatically crawling related bug reports from open-source repositories, which serve as the knowledge base for three LLM agents in \ourtool. These agents operate in a multi-agent fashion as follows:
\begin{itemize}[leftmargin=*]
    \item \textbf{Mutator Invention Agent:} Analyzes historical bug reports to identify error-prone language features, such as those frequently associated with compiler bugs, and proposes mutator designs as specifications describing the intended transformations \bocomment{(Section~\ref{sec:approach:invention})}.
    \item \textbf{Mutator Implementation Synthesis Agent:} Synthesizes concrete mutator code guided by the specifications and high-quality few-shot mutator examples that illustrate AST modification code \bocomment{(Section~\ref{sec:approach:impl})}.
    It is also fine-tuned with a curated set of \bocomment{human-written} seed mutators to produce more accurate and diverse transformations \bocomment{(Section~\ref{sec:approach:sft})}.
    \item \textbf{Mutator Refinement Agent:} Validates each synthesized mutator by applying it to bug-triggering programs collected from historical bug reports \bocomment{(Section~\ref{sec:approach:refine})}. If compilation or runtime errors occur, the agent automatically repairs them using error feedback. This process iterates until validation succeeds or a retry limit is reached.
\end{itemize}

Following this workflow, we apply \ourtool to two prominent system-level programming languages, Rust and C++, and automatically analyze 1,000 bug reports (500 each for Rust and C++) from two open-source repositories: rustc for Rust and Clang for C++.
\bocomment{To guide the LLM toward more accurate and up-to-date use of AST modification APIs}, we fine-tune the LLM of the last two agents using a small set of human-written mutators as demonstrations.
\bocomment{In our main evaluation}, \ourtool synthesizes 319 mutators for Rust and 403 mutators for C++. 
\bocomment{The end-to-end API cost is approximately \$0.08 per final valid mutator using GPT-4o, where the cost includes invention, implementation synthesis, and refinement, but excludes the one-time fine-tuning cost.}
\bocomment{To study the impact of the bug-report budget, we further analyze 200 additional bug reports per language, contributing 125 and 158 additional valid mutators for Rust and C++, respectively.}
These operators target complex and error-prone language features, such as trait object layouts in Rust\footnote{\url{https://doc.rust-lang.org/book/ch10-02-traits.html}} and templates in C++\footnote{\url{https://en.cppreference.com/w/cpp/language/templates.html}}, which remain underexplored in prior work~\cite{even2023grayc,ou2024mutators,wang2025fuzzing,sharma2023rustsmith,wang2024rustlantis,yang2024rust}.

The synthesized mutators are then integrated into a customized fuzzer targeting production-level compilers, including rustc, gccrs, GCC, and Clang. 
Seeded with the enhanced seed program set, \ourtool uses crash and hang oracles, as well as differential testing across compilers of the same language (e.g., GCC and Clang for C++).
We benchmark \ourtool against a range of state-of-the-art compiler fuzzers, including traditional fuzzers such as RustSmith~\cite{sharma2023rustsmith} and Rustlantis~\cite{wang2024rustlantis} for Rust, and CSmith~\cite{yang2011finding}, YARPGen~\cite{livinskii2020random,livinskii2023fuzzing}, GrayC~\cite{even2023grayc}, 
and TyMut~\cite{wang2025fuzzing} for C++.
Additionally, we compare against \textit{the recent LLM-based fuzzers}, including Clozemaster~\cite{gao2025clozemaster} for Rust, as well as Fuzz4All~\cite{xia2024fuzz4all} and MetaMut~\cite{ou2024mutators} for C++.
The results indicate that \ourtool outperforms existing approaches by discovering more unique crashes, achieving the highest line coverage on Rust and both C++ compilers.
Particularly, compared with the approaches that directly leverage LLMs to mutate code (i.e., Clozemaster~\cite{gao2025clozemaster} and Fuzz4All~\cite{xia2024fuzz4all}), our approach consistently performs better.
\bocomment{Additional ablation and sensitivity studies further validate the necessity of the key components and the robustness of the default configurations of \ourtool.}

In the bug-hunting campaign \bocomment{based on 700 bug reports per language}, \ourtool uncovered \bocomment{100} bugs across both languages.
For the Rust compilers (rustc and gccrs), \ourtool reported 62 bugs, of which \bocomment{44} were confirmed as previously unknown.
For the C++ compilers (GCC and Clang), \ourtool reported \bocomment{38} bugs, including 17 confirmed as new.
\bocomment{Overall, 61 bugs have been confirmed as previously unknown, and 35 of them have already been fixed by developers.}
These results demonstrate that \ourtool can effectively synthesize high-quality mutators for complex language features, enabling practical bug discovery in real-world compilers.

\smalltitle{Contributions}
\ourtool addresses the key limitations of existing mutation-based compiler fuzzers.
To overcome the limited scope and expressiveness of existing mutators \textit{(C1)}, it automatically invents new mutators guided by compiler-specific insights from open-source bug reports, often covering advanced language features. Unlike prior approaches that require manual effort to design mutators \textit{(C2)} and are typically restricted to a single language or compiler \textit{(C3)}, \ourtool leverages LLMs as the generation engine—minimizing human intervention and enabling cross-language transfer through LLMs' multilingual knowledge. In summary, this work makes the following contributions:
\begin{itemize}[leftmargin=*]
    
    \item We propose \ourtool, a multi-agent framework that synthesizes high-quality mutators for production-level compilers, guided by insights from historical compiler bug reports widely available on open-source communities.

    \item We propose fine-tuning LLMs to guide the generation \bocomment{toward more accurate and up-to-date} APIs within mutators, and introduce an adaptive seed enhancement strategy to diversify the seed set, thereby improving the effectiveness of mutation-based compiler fuzzing.

    \item \ourtool synthesizes 722 executable mutators (319 for Rust, 403 for C++), each costing approximately \$0.08 using GPT-4o APIs. \bocomment{Additional bug-report budget analysis further shows that more reports can synthesize more valid mutators and improve compiler coverage.}

    \item \bocomment{We demonstrate the effectiveness of \ourtool by reporting 100 compiler bugs across rustc, gccrs, GCC, and Clang, 61 of which have been confirmed as previously unknown, with 35 already fixed by developers.}
    Additionally, Mut4All outperforms all baselines by achieving the highest coverage on Rust and C++ compilers and \bocomment{recovering 117 exclusive crashes that all other fuzzers missed, including 104 in Rust compilers and 13 in C++ compilers}.
\end{itemize}


%% file: Motivating.tex
\section{Illustrative Examples}
\label{sec:example}

To demonstrate \ourtool's ability to generate sophisticated, feature-specific mutators, we present two confirmed bugs it uncovered in rustc and GCC.

\subsection{An Example for Rust}


\begin{figure}[t]
  \centering
  \begin{subfigure}[t]{0.48\linewidth}
    \centering
    \includegraphics[width=\linewidth]{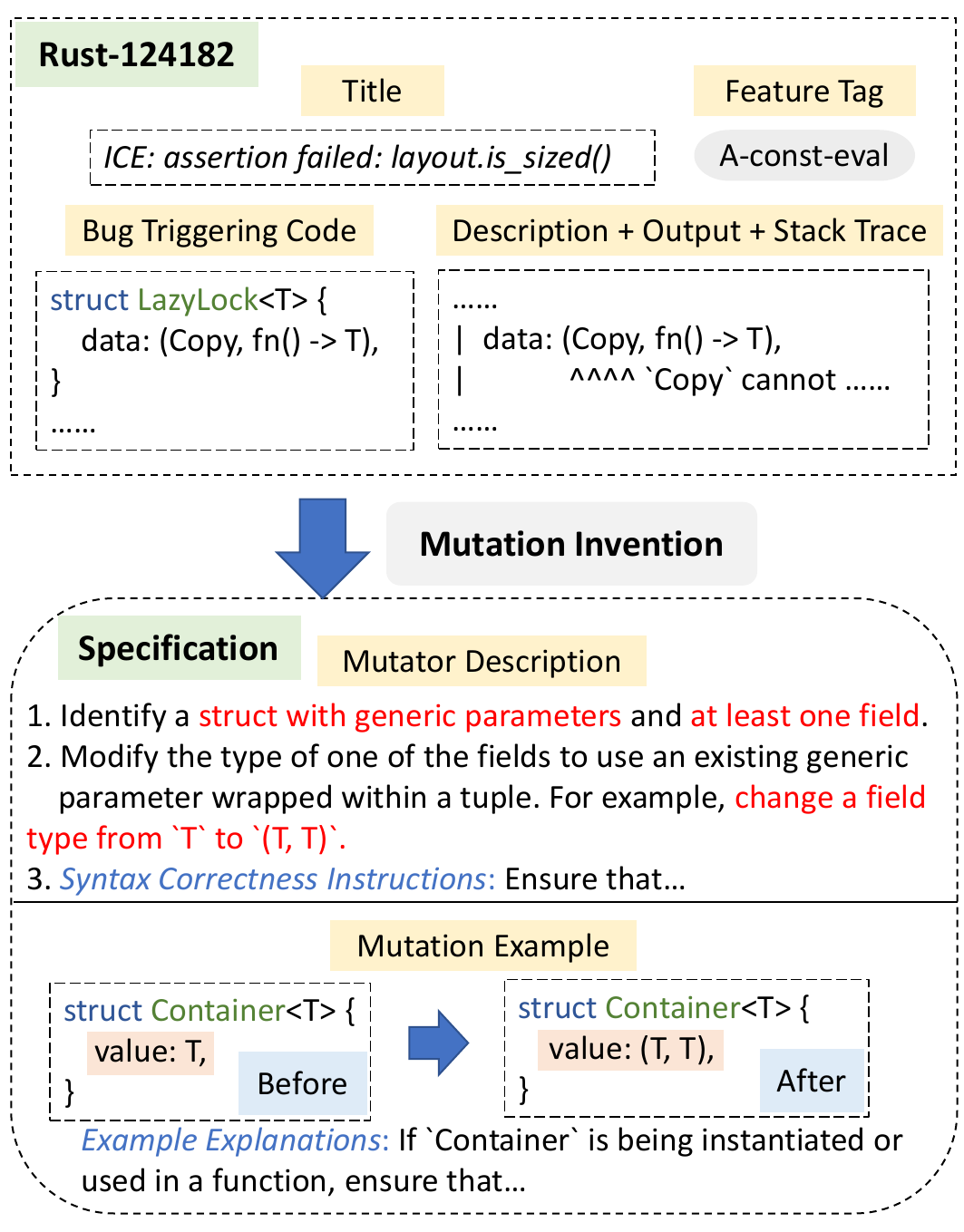}
    \caption{Rust mutator specification via Rust-124182.}
    \figlabel{example-invention-rust}
  \end{subfigure}
  \quad
  \begin{subfigure}[t]{0.48\linewidth}
    \centering
    \includegraphics[width=\linewidth]{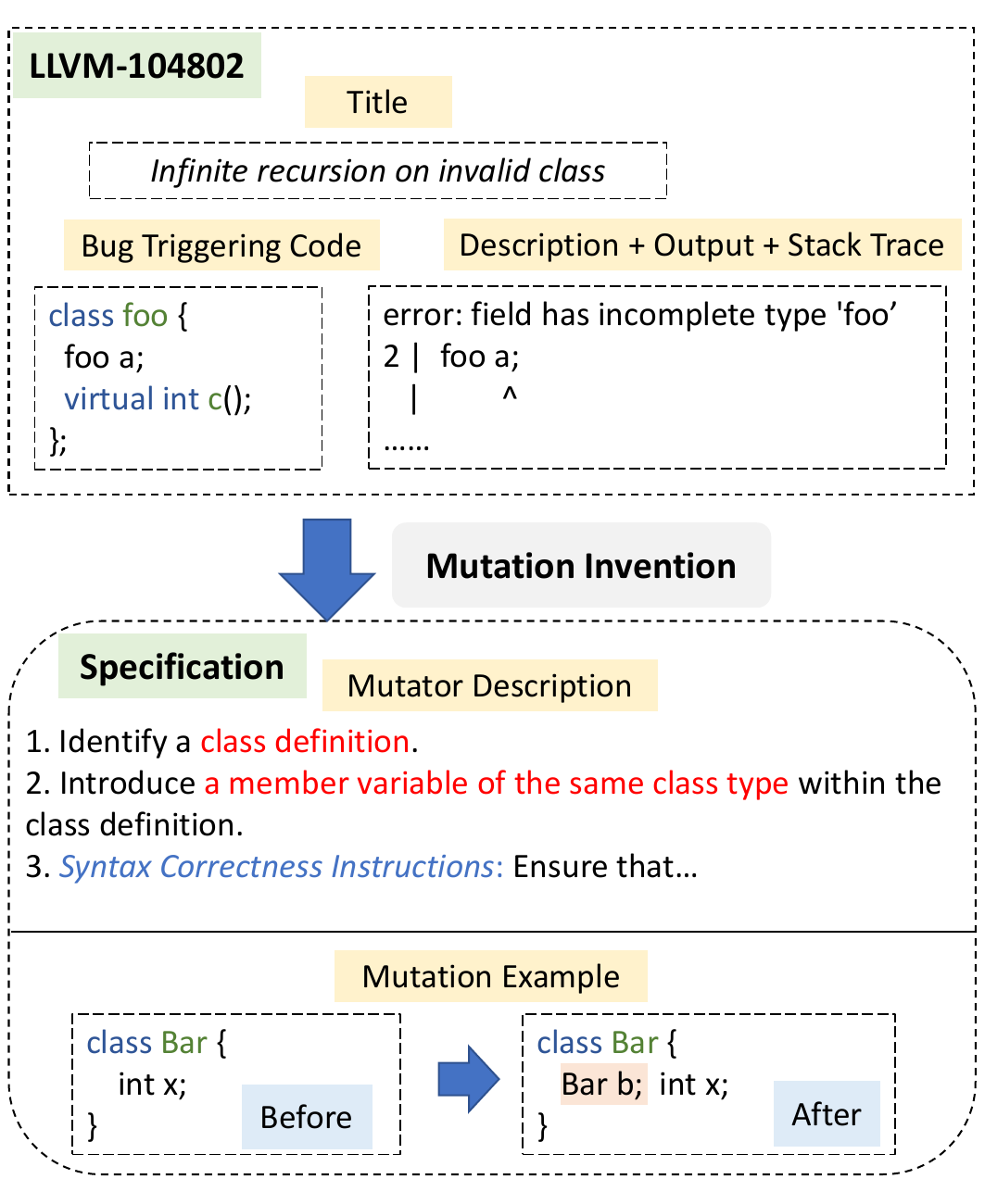}
    \caption{C++ mutator specification via LLVM-104802.}
    \figlabel{example-invention-Cpp}
  \end{subfigure}

  \caption{Mutator specification examples generated by \ourtool.}
  \figlabel{example-invention}
\end{figure}

We first demonstrate how \ourtool generates mutators for Rust by showcasing a confirmed bug it uncovered in the official Rust compiler (rustc). The process begins by collecting historical bug reports from the Rust compiler's official GitHub repository, each typically including a descriptive title, a triggering code snippet, compiler output, and a stack trace, as shown in the top half of \figref{example-invention-rust}. These reports provide valuable insights into the compiler's behavior and highlight language features that are insufficiently tested and may lead to new bugs.

Given this bug report, \ourtool invokes a pipeline of three LLM agents to generate a new mutator. The first agent, \textit{Mutator Invention}, analyzes the bug description and the triggering code to identify key language features responsible for the bug. It then synthesizes a mutator specification comprising a detailed description, syntax well-formedness constraints, and a concrete \textit{Before} and \textit{After} example illustrating the transformation. As shown in \figref{example-invention-rust}, the historical bug (Rust-124182\footnote{\url{https://github.com/rust-lang/rust/issues/124182}}) is recognized by \ourtool as related to generic parameter handling and type layout, which is based on the presence of a generic type in the code snippet and a reference to type layout in the bug title. This insight guides the LLM agent to invent a mutator that duplicates a generic type and wraps it inside a tuple, as detailed in the specification in the bottom half of \figref{example-invention-rust}.
Note that this mutator \bocomment{targets a more feature-specific transformation} than those generated by existing methods such as MetaMut~\cite{ou2024mutators}, which typically rely on a fixed set of syntactic transformations.
In contrast, our approach leverages insights extracted from real-world bug reports to design mutators that \bocomment{cover language-specific transformations less directly captured by these existing techniques}.

\begin{codelist}[t]
\begin{lstlisting}[style=mystyleRust,caption={The core implementation of the Rust mutator derived from Rust-124182.}, captionpos=b, label=lst:rustMutatorExample]
//  Match generic type `T`
if let Type::Path(TypePath { path, .. }) = &first_field.ty {
    if let Some(segment) = path.segments.first() {
        if item_struct.generics.params.iter().any(|param| {
            if let GenericParam::Type(type_param) = param {
                type_param.ident == segment.ident
            } else {
                false
            }
        }) {
            // Duplicate `T` and replace the type with `(T, T)`
            let generic_ident = segment.ident.clone();
            first_field.ty = parse_quote!((#generic_ident, #generic_ident));
        }
    }
}
\end{lstlisting}
\end{codelist}

\begin{codelist}[t]
\begin{lstlisting}[style=mystyleRust,caption={The new bug Rust-141911 uncovered by Mut4All.}, captionpos=b, label=lst:rustNew]
trait MyTrait { fn virtualize(&self); }
struct VirtualWrapper<T>(T, T); // From `(T)` to `(T, T)`
impl<T: 'static> MyTrait for T {
    fn virtualize(&self) {
        const { std::ptr::null::<VirtualWrapper<T>>() as *const dyn MyTrait };
    }
}
fn main() { 0u8.virtualize(); }
\end{lstlisting}
\end{codelist}

Next, the \textit{Mutator Implementation Synthesis} agent leverages the generated specification, our customized prompt templates, and several AST transformation examples extracted from Rust's official AST library to produce a concrete mutator that performs the intended program transformation.
The backend LLM of this agent is already enhanced via fine-tuning with a set of high-quality human-written mutators across all target languages.
The core logic of the resulting mutator is shown in Listing~\ref{lst:rustMutatorExample}, which matches a generic type \code{T} at the AST level and replaces it with a tuple type \code{(T, T)}. 
\bocomment{This transformation targets compiler logic related to generic type handling and type layout, reflecting the failure pattern observed in the original bug report.}
In this case, the LLM generates a syntactically correct and semantically valid mutator without requiring any repair iterations.
As a result, the \textit{Mutator Refinement} agent is skipped for this example.

During our fuzzing phase, the mutator generates new program variants that stress-test the compiler's type system in unexpected ways. As shown in Listing~\ref{lst:rustNew}, Line 2 is mutated by introducing a self-referential tuple type \code{VirtualWrapper<T>(T, T)}. This transformation causes the compiler to cast a pointer to the trait object \code{dyn MyTrait} into a pointer to the concrete type \code{VirtualWrapper<T>}, which is disallowed. The compiler fails to handle this edge case properly, resulting in an internal assertion failure. This bug, reported as Rust-141911, has been confirmed by Rust compiler developers as a previously unknown crash in rustc.

\subsection{An Example for C++}
Similarly, \ourtool generates mutators for C++ compilers by analyzing bug reports from LLVM's GitHub repository. The process mirrors that of Rust, where LLM agents collaboratively invent, implement, and refine mutators based on compiler-specific insights. \figref{example-invention-Cpp} presents an example produced by \ourtool for C++.

The original bug report (LLVM-104802\footnote{\url{https://github.com/llvm/llvm-project/issues/104802}}) describes a crash in the Clang C++ compiler triggered by a class that declares a member of its own type. In this case, the compiler fails to handle the invalid structure gracefully and enters infinite recursion during type checking. This suggests that recursive type patterns are not robustly guarded in certain compilation paths. Based on this insight, \ourtool's \textit{Mutator Invention} agent designs a mutator that intentionally injects self-referential type declarations, with the corresponding mutator specification shown in the bottom half of \figref{example-invention-Cpp}. As mentioned above, existing methods such as MetaMut~\cite{ou2024mutators} are unable to design such mutators, as they rely on limited predefined syntactic transformations and do not account for language-specific features such as self-referential types.

Next, the \textit{Mutator Implementation Synthesis} agent generates an implementation that inserts a self-referential field into a struct. Initially, the LLM produces a mutator using a deprecated AST API, resulting in a compiler error. After receiving the error feedback, \ourtool's \textit{Mutator Refinement} agent successfully repairs the mutator and produces a correct version, shown as Listing~\ref{lst:cppMutatorExample}.
The final mutator introduces recursive types, such as introducing a field of type \code{S} (i.e., \code{S t;}) into the struct, provoking recursive type behavior and potentially unsafe casting.
Listing~\ref{lst:cppNew} shows an example of the mutated program, where the inserted self-reference interferes with a bit cast operation. This results in a compiler crash, reported as bug GCC-120798, which has been confirmed by GCC maintainers.

\begin{codelist}[t]
\begin{lstlisting}[style=mystyleCpp,caption={The core implementation of the C++ mutator derived from LLVM-104802.}, captionpos=b, label=lst:cppMutatorExample]
void MutatorFrontendAction_6::Callback::run(const MatchFinder::MatchResult &Result) {
    //Check whether the matched AST node is the target node
    if (auto *CD = Result.Nodes.getNodeAs<clang::CXXRecordDecl>("ClassDecl")) {
      //Filter nodes in header files
      if (!CD || !Result.Context->getSourceManager().isWrittenInMainFile(CD->getLocation()))  
          return;
      if (!CD->isCompleteDefinition())  
          return;
      //Get the source code text of target node
      auto className = CD->getNameAsString();
      if (className.empty())  
          return;
      //Perform mutation on the source code text by applying string replacement
      auto classText = stringutils::rangetoStr(*(Result.SourceManager), CD->getSourceRange());
      std::string memberDecl = className + " self_member;";
      memberDecl = "\n/*mut6*/" + memberDecl;
      classText = classText.insert(classText.rfind("}")-1, memberDecl);
      //Replace the original AST node with the mutated one
      Rewrite.ReplaceText(CharSourceRange::getTokenRange(CD->getSourceRange()), classText);
    }
}
\end{lstlisting}
\end{codelist}

\begin{codelist}[t]
\begin{lstlisting}[style=mystyleCpp,caption={The new bug GCC-120798 uncovered by Mut4All.}, captionpos=b, label=lst:cppNew]
struct S { 
  int s;
  S t; // Inserted by Mut4All
};
auto t = __builtin_bit_cast (S, (0)); 
\end{lstlisting}
\end{codelist}

%% file: Approach.tex
\section{Approach}
\label{sec:approach}

\begin{figure*}[t!]
    \centering
    \includegraphics[width=\linewidth]{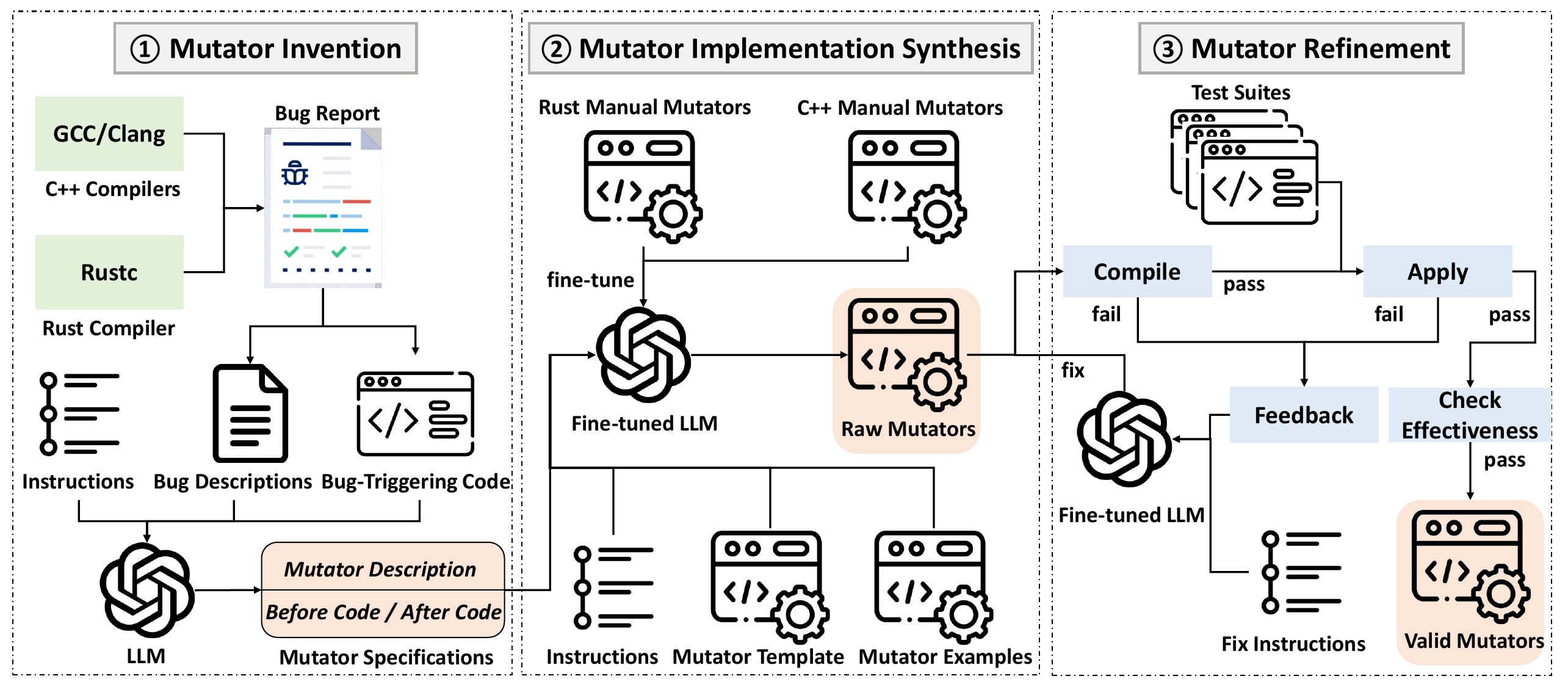}
    \caption{The overview of the interaction of the three \ourtool agents.}
    \figlabel{overview}
\end{figure*}

We now describe the design of \ourtool, with a detailed explanation of the processes underlying each step shown in the examples in Section~\ref{sec:example}.
\figref{overview} shows an overview of \ourtool agents, briefly describing the workflow and detailing the inputs and outputs of each component. Overall, \ourtool takes as input historical bug reports, prompt templates for each LLM agent, and existing manually written mutator examples, and produces a set of validated mutators ready for integration into our subsequent compiler fuzzing phase.
At a finer granularity, as mentioned above, \ourtool comprises three specialized LLM agents corresponding to three key steps:
(1) the \textit{Mutator Invention} agent, which takes bug reports and bug-triggering code snippets as input and outputs a set of \textit{mutator specifications};
(2) the \textit{Mutator Implementation Synthesis} agent, which takes these specifications, along with our prompt templates and human-written mutator examples, to implement raw mutators in the target language; and
(3) the \textit{Mutator Refinement} agent, which takes the raw mutators and their associated compiler error messages (if any) as input and outputs validated mutators ready for large-scale compiler fuzzing.
\bocomment{To reduce ambiguity and hallucination in interactions with LLMs, we design the prompts in a task-specific and structured manner. For each agent, the prompt explicitly specifies the role, task, functional instructions, output format, and required inputs, so that the LLM only focuses on one well-defined subtask at a time.}
We now describe the design of each agent in detail.

\subsection{Mutator Invention} \label{sec:approach:invention}

Compilers, as the foundation of modern software development, are among the most actively maintained and scrutinized components in the open-source ecosystem. Bugs and usage issues are frequently reported on platforms such as GitHub, often accompanied by detailed discussions, triggering code snippets, and diagnostic outputs. These bug reports reflect valuable domain knowledge from compiler developers and often highlight specific language constructs or internal behaviors prone to failure. \ourtool leverages these historical bug reports as input to its first component, the \textit{Mutator Invention} agent. This agent analyzes the bug descriptions and associated code to identify underlying language features responsible for the reported failures, and uses this analysis to design targeted mutation operators. It is based on the observation that many historical compiler bugs arise from recurring constructs or subtle corner cases that are difficult to expose via mutators that are supported by existing approaches. By grounding the mutation logic in real-world failure scenarios, \ourtool can generate semantics-aware, feature-specific mutators that are more likely to trigger previously unseen bugs.

To support this process, we design a structured prompt template (shown in Figure~\ref{invention-temp}) that guides the LLM to extract relevant language features from each bug report and synthesize a precise, context-aware mutation operator. The prompt is designed with clear functional instructions: it explicitly requires the LLM to avoid superficial edits, target well-defined code constructs, and design mutations that achieve high coverage across diverse code contexts. To improve generalizability, it discourages the use of ad hoc identifiers and promotes the reuse of existing code elements.
\bocomment{Although this agent uses a single prompt, it targets one well-defined task, namely generating exactly one mutator specification from one bug report. The intermediate steps in the prompt serve only as structured guidance toward this output.}

\begin{figure}[t!]
\begin{tcolorbox}[boxrule=1pt, left=4pt, right=4pt, top=2pt, bottom=2pt]

\textbf{[Role]}: You are a \textbf{\{Language\}} compiler expert.

\textbf{[Task]}: Here is a bug report for the \textbf{\{Language\}} compiler labeled by \textbf{\{Tag\}}.
Summarize the structures that \textbf{\{Language\}} compiler bugs are more likely to trigger, and design mutation operators...

\textbf{[Functional Instructions]}: 1. Target a specific code construct and clearly state where it applies; 2. Customized data names, types, and variable names...; 3. Achieve high coverage across different program structures and features; 4. ...

\textbf{[Output Instructions]}: The output must be plain text. Do not use the Markdown format. Design exactly one mutation operator per request...

\textbf{[Input]}: Here are the contents of the bug report: \textbf{\{Bug Descriptions\}} and \textbf{\{Bug-Triggering Code\}}.
\end{tcolorbox}
\caption{Prompt template for mutator invention.}
\label{invention-temp}
\end{figure}

\subsection{Mutator Implementation Synthesis} \label{sec:approach:impl}

Once a mutator has been invented, the next step is to synthesize an executable implementation. This typically involves writing AST transformation code that modifies a program according to the intended mutation—a process that is tedious and error-prone when done manually, especially for complex and evolving compiler infrastructures. To address this, \ourtool introduces a second agent responsible for \textit{Mutator Implementation Synthesis}. This LLM-based agent generates transformation code tailored to the target language based on the mutator specification. \bocomment{To improve code quality and reduce API-related errors}, we fine-tune the LLM using OpenAI's standard fine-tuning API, guided by 10 human-written mutator examples per language. 
These examples cover diverse syntactic constructs and representative transformation patterns, serving as domain-specific demonstrations that help the model synthesize \bocomment{more} accurate and idiomatic AST manipulation code.

To enforce structural consistency, \ourtool adopts a predefined mutator template for each supported language. The template defines the skeleton that the LLM must complete with appropriate mutation logic like matching AST nodes, applying transformations, and handling edge cases.
Listing~\ref{lst:rustTemp} and Listing~\ref{lst:cppTemp} present simplified examples for Rust and C++, respectively.
\bocomment{Concretely, for Rust we implement mutators on top of the \code{syn}\footnote{\url{https://docs.rs/syn/latest/syn/}} framework, while for C++ we build on Clang LibTooling\footnote{\url{https://clang.llvm.org/docs/LibASTMatchersTutorial.html}} (e.g., \code{MatchFinder}, \code{ASTContext}, and \code{Rewriter}) for AST analysis and transformation.}
Each template specifies the entry point of one mutator, filtering conditions, and placeholders for inserting LLM-generated transformation code.
These mutator templates \bocomment{are manually designed by us as language-specific scaffolds following standard AST matching and transformation workflows.} They are general and expressive enough to capture the full operational logic of all mutators, thereby providing a unified structure for matching and transforming code.

As illustrated in \figref{overview}, the Mutator Implementation Synthesis agent combines three key ingredients: (1) the mutator specification, (2) the language-specific mutator template, and (3) human-crafted AST transformation examples.
It invokes the fine-tuned LLM with a structured prompt (shown in Figure~\ref{mutgen-temp}) that clearly defines the agent's role, functional constraints (e.g., correct API usage, no syntax errors), and output format. The output of this agent is a raw mutator implementation, which is subsequently passed to the \textit{Mutator Refinement} agent for potential repair and validation.

\begin{figure}[t]
\begin{tcolorbox}[boxrule=1pt, left=4pt, right=4pt, top=2pt, bottom=2pt]
\textbf{[Role]}: You are an expert in \textbf{\{Language\}} compiler development and AST manipulation.

\textbf{[Task]}: Based on the provided mutator specification and example mutators for the target language, complete the \textbf{\{Language\}} mutator template to implement the intended transformation.

\textbf{[Functional Instructions]}: 1. Ensure that the generated code is syntactically valid \textbf{\{Language\}}; 2. Correctly implement the mutation logic as described in the specification; 3. Follow the structure of the provided template; 4. Use appropriate \textbf{\{Language\}} APIs for AST manipulation; 5. Include all necessary imports and dependencies...

\textbf{[Output Instructions]}: Output only the completed \textbf{\{Language\}} code, without any additional explanations or formatting...

\textbf{[Input]}: Here are the necessary inputs:

Mutator Specification: \textbf{\{Mutator Specification\}}.

Mutator Template for \textbf{\{Language\}} : \textbf{\{Mutator Template\}}.

Human-written \textbf{\{Language\}} mutators: \textbf{\{Mutator Examples\}}.
\end{tcolorbox}
\caption{Prompt template for mutator implementation synthesis.}
\label{mutgen-temp}
\end{figure}

\begin{codelist}[t]
\begin{lstlisting}[style=mystyleRust,caption={The core part of the mutator template for Rust.}, captionpos=b, label=lst:rustTemp]
impl Mutator for Mutator_ {
    // The name of the mutator
    fn name(&self) -> &str {
        "Mutator_"
    }
    // The function for traversing AST nodes and mutating.
    fn mutate(&self, file: &mut syn::File) {
        for item in &mut file.items {
            // The details of the implementation of mutators
        }
    }
}
\end{lstlisting}
\end{codelist}


\begin{codelist}[t]
\begin{lstlisting}[style=mystyleCpp,caption={The core part of the mutator template for C++.}, captionpos=b, label=lst:cppTemp]
// Mutating the target AST node (i.e., insertion, deletion, replacement) 
void MutatorFrontendAction_{{bugno}}::Callback::run(const MatchFinder::MatchResult &Result) {
    if (auto *MT = Result.Nodes.getNodeAs<clang::{{ASTNodeType}}>({{NodeName}})) {
      if (!MT || !Result.Context->getSourceManager().isWrittenInMainFile(MT->getLocation()))
        return;
    }
}
// Localizing the target AST node, and binding the mutation operations
void MutatorFrontendAction_{{bugno}}::MutatorASTConsumer_{{bugno}}::HandleTranslationUnit(ASTContext &Context) {
    MatchFinder matchFinder;
    DeclarationMatcher matcher = {{MatchRule}}.bind({{NodeName}});
    Callback callback(TheRewriter);
    matchFinder.addMatcher(matcher, &callback);
    matchFinder.matchAST(Context);
}
\end{lstlisting}
\end{codelist}

\subsection{Mutator Refinement}  \label{sec:approach:refine}

In this agent, each raw mutator generated in the above stage undergoes validation and potential repair by the \textit{Mutator Refinement} agent. \ourtool begins by extracting all bug-triggering programs from the original bug reports and assembling them into a seed test suite. These programs are used to validate whether the mutator (1) compiles successfully, (2) integrates with the compiler's AST infrastructure, and (3) performs its transformation without introducing compilation failures in downstream programs.

If one mutator fails to compile, or if applying it to any seed program causes a compilation error, \ourtool captures the full compiler error messages (e.g., due to missing imports, incorrect type handling, or invalid AST rewrites). These messages are embedded into a structured prompt template (Figure~\ref{refinement-temp}) that instructs the LLM to revise the mutator while preserving its intended transformation. The refinement process is conducted in an iterative loop. After each repair attempt, the revised mutator is recompiled and re-applied across all seed programs.
If the mutator implementation successfully compiles and does not cause any errors when applied to the test suite, it is marked as valid and added to the mutator pool for fuzzing. Otherwise, the process continues for up to a maximum of $N$ iterations (we use $N=10$ in our experiments). If the mutator still fails validation after the limit is reached, it is discarded.

Furthermore, to ensure that the mutator performs meaningful transformations, \bocomment{we include an additional check by applying each compiled mutator to the prepared validation test suite. If the mutator runs successfully but does not modify any file in this test suite at all (e.g., due to overly strict matching logic), it is considered ineffective and discarded. In other words, our current criterion is binary: a mutator is treated as ineffective if it produces zero syntactic changes across all validation inputs.}
This ensures that the final mutator set is not only valid, but actively contributes to provide meaningful and diverse mutations for the fuzzing phase.

\begin{figure}[t]
\begin{tcolorbox}[boxrule=1pt, left=4pt, right=4pt, top=2pt, bottom=2pt]

\textbf{[Role]}: You are a \textbf{\{Language\}} compiler expert revising a mutator code to fix errors while preserving its mutation logic.

\textbf{[Task]}: Given the buggy mutator code and the associated compiler error messages, produce a corrected version that resolves the errors without changing the core mutation behavior.

\textbf{[Functional Instructions]}: 1. Prioritize fixing the provided compiler errors by identifying their causes (e.g., missing imports, type mismatches) and applying targeted corrections. 2. Ensure the revised code is syntactically valid \textbf{\{Language\}} and compiles successfully. 3. ...

\textbf{[Output Instructions]}: Output only the corrected \textbf{\{Language\}} code in plain text, without any explanations, comments, or Markdown formatting.

\textbf{[Input]}: Here are the necessary inputs:

Previous Mutator Code: \textbf{\{Previous Code\}}

Compile Errors: \textbf{\{Compile Errors\}}

\end{tcolorbox}
\caption{Prompt template for mutator refinement.}
\label{refinement-temp}
\end{figure}

\subsection{Fine-Tuning} \label{sec:approach:sft}
Due to the frequent evolution of compiler infrastructures and their underlying AST manipulation libraries, many APIs encoded in the pretrained LLM’s knowledge have become outdated or deprecated across compiler versions, which can lead to incorrect or obsolete API usage when synthesizing mutators. To mitigate this issue, we fine-tune the Mutator Implementation Synthesis agent on a set of up-to-date, manually designed mutator specifications, \bocomment{helping it better follow} the canonical and version-consistent use of modern AST APIs. Specifically, we manually construct 10 mutator specifications for each language as demonstration data.
For Rust, we construct 10 mutator specifications and their corresponding implementations derived from representative bug reports; for C++, we build upon 10 mutators from our prior work, TyMut~\cite{wang2025fuzzing}, and refine both their specifications and implementations based on real bug reports.
\bocomment{We intentionally do not hard-code a language-specific API whitelist into the prompts, as doing so would make the framework more tightly coupled to particular compiler infrastructures and versions, thereby reducing its cross-language adaptability. Instead, we rely on fine-tuning to reduce API-related errors while preserving a more general prompting strategy.}
These carefully designed mutator–specification pairs jointly serve as supervised examples for fine-tuning, covering diverse AST manipulation patterns and language-specific features observed in real compiler ecosystems.

\bocomment{These examples cover all six categories in the mutator taxonomy discussed in Section~\ref{sec:taxonomy-valid-mutators}, including declaration-level, type-structure, generic / template-related, constraint / bound-related, expression / attribute-level, and layout / cast / initialization-related mutations.}

We fine-tune GPT using Supervised Fine-Tuning (SFT), where a token-level cross-entropy loss based on maximum likelihood estimation is applied to minimize the discrepancy between generated text and human demonstrations. The loss function is defined as:
\begin{equation}
\mathcal{L}_{\text{SFT}}(\theta) = - \sum_{t=1}^{T} \log P_{\theta}(y_t \mid y_{<t}, x)
\end{equation}
where $x$ denotes the input prompt, $y = (y_1, y_2, \dots, y_T)$ is the target sequence of human demonstrations with length $T$. $P_{\theta}(\cdot)$ represents the conditional probability distribution parameterized by the model parameters $\theta$, and $y_{<t}$ denotes the prefix of tokens before the current time step $t$. In our case, the human demonstrations correspond to manually constructed mutator–specification pairs, where each specification serves as the input prompt and its corresponding mutator implementation serves as the target output.
By minimizing this loss, the model is trained to maximize the likelihood of producing the human-written target sequence, \bocomment{thus encouraging its generation behavior to better follow up-to-date, version-stable compiler APIs. We do not use a separate validation split because fine-tuning is used only as lightweight guidance for AST API usage. 
Its effectiveness is evaluated on separate bug reports through downstream mutator synthesis, compilation-based validation, and iterative refinement.}

\subsection{The Main Fuzzing Procedure}
\label{sec:main-fuzzing}

\begin{algorithm}[tp]
\small
\caption{The Fuzzing Procedure of \ourtool.}
\label{alg:loop}

\KwIn{Seed Programs: $S$}
\KwIn{Valid Mutators: $M$}
\KwIn{Compilers Under Test for a Certain Language: $C$}
\KwIn{Timeout: $T$}

\BlankLine

\SetKwFunction{Enhance}{Enhance}
\SetKwFunction{Expired}{Expired}
\SetKwFunction{RandomPick}{RandomPick}
\SetKwFunction{Apply}{Apply}
\SetKwFunction{Compile}{Compile}
\SetKwFunction{Crash}{Crash}
\SetKwFunction{Hang}{Hang}
\SetKwFunction{Record}{Record}
\SetKwFunction{Execute}{Execute}
\SetKwFunction{Inconsistent}{Inconsistent}

$\mathcal{P} \leftarrow\Enhance{S}$ \\


\While{$\neg \Expired{T}$}{

    \For{$i \leftarrow 1$ \KwTo $\operatorname{Int.Max}$}{
        $M' \leftarrow \RandomPick{M, i} $ \\
        \ForEach{$p \in \mathcal{P}$}{
            \ForEach{$m \in M'$}{
                $p' \leftarrow \Apply{p,m}$ \\
                $R \leftarrow \phi$ \\
                \ForEach{$c \in C$}{
                    ${res, exe} \leftarrow \Compile{c, p'}$ \\
                    \If{$\Crash{res} ~ \operatorname{OR} ~  \Hang{res}$ } {
                        $\Record{p', c}$ \\
                    }
                    \ElseIf{$\{ {exe} \} \neq \phi$} {
                        $r \leftarrow \Execute{exe}$ \\
                        $R \leftarrow R \cup \{ r \}$ \\
                    }
                }
                \If{$\Inconsistent{R}$} {
                    $\Record{p', R}$ \\
                }
            }
        }
    }
}
\end{algorithm}

After collecting a set of validated mutators, \ourtool enters its core fuzzing loop to test target compilers systematically. Algorithm~\ref{alg:loop} outlines this process in detail.

Given an initial set of seed programs $S$, \ourtool first expands the seed corpus using rule-based transformations via the \textit{enhance} function (introduced in detail in Section~\ref{sec:enhancement}).
This step aims to diversify the seed space and uncover additional transformation opportunities, which will be elaborated on later. The enhanced seed corpus is denoted as $\mathcal{P}$.
During the fuzzing loop, in each iteration, \ourtool randomly selects $i$ mutators from the validated mutator set $M$ and applies them to each program in the current seed pool $\mathcal{P}$. The resulting mutated programs $p'$ are then compiled and executed on each target compiler $c$ in the compiler set $C$. 
\bocomment{Note that \emph{one mutation operator per request} refers only to the invention stage. During the fuzzing loop, multiple synthesized mutators can be applied to a seed program.}
If a compiler crashes or hangs, the corresponding input is immediately recorded as a bug-triggering example. For successfully compiled programs, the execution results $r$ of the generated binaries are collected. \ourtool compares these outputs across compilers or against expected values (when available). If any inconsistency is detected—such as divergent outputs, abnormal termination, or unexpected side effects---a behavioral bug is recorded. The fuzzing loop continues until a predefined timeout $T$ is reached or until other stopping criteria (e.g., maximum number of iterations or detected bugs) are met.
\bocomment{These oracles can expose compiler failures and observable inconsistencies, including some bugs in frontend semantic-analysis components, but they do not constitute a formal oracle for proving language-level soundness violations.}
For all recorded programs, we manually inspected them for potential undefined behavior and cross-checked existing bug reports to determine whether they are duplicates before submitting new issues to the open-source compiler communities.

\subsection{Adaptive Search-Based Seed Enhancement}
\label{sec:enhancement}

\subsubsection{Algorithm}

\begin{algorithm}[t]
\DontPrintSemicolon
\small
\caption{Adaptive Seed Enhancement.}
\label{alg:type-guided-mutation}
\KwIn{Seed set of programs $\mathcal{P}$} 
\KwIn{Retry limit $T$}
\KwIn{The lower bound of weight $w_{lb} \in [0,1]$}
\KwIn{Decay factor $\gamma \in (0,1)$}
\KwOut{Mutated seed program set $\mathcal{P}'$}

\SetKwFunction{Mutate}{Mutate}
\SetKwFunction{Build}{BuildTypeIndex}
\SetKwFunction{Select}{SelectSameType}
\SetKwFunction{Check}{Check}
\SetKwFunction{Replace}{Replace}
\SetKwFunction{TypeOf}{TypeOf}
\SetKwFunction{Len}{Len}
\SetKwFunction{Rand}{Random}
\SetKwFunction{WChoice}{WeightedChoice}
\SetKwFunction{Min}{Min}
\SetKwFunction{Max}{Max}

\BlankLine

$\mathcal{P}' \leftarrow \mathcal{P}$ \tcp*[r]{initialize the output set}

$\mathsf{Index} \leftarrow \emptyset$ \tcp*[r]{$\mathsf{Index}: \text{AST type} \mapsto \{\text{subtrees}\}$}

\ForEach{$\mathsf{prog} \in P$}{
    \ForEach{$n$ in $\mathsf{prog}$}{
        $\mathsf{Index}[\TypeOf{n}]  \leftarrow \mathsf{Index}[\TypeOf{n}] \cup \{n\}$ \\
        $w(n) \leftarrow w_{lb} + (1-w_{lb})/$\Len{$n$} \tcp*[r]{initialize weights}
    }
}
\BlankLine

$\mathsf{Used} \leftarrow \emptyset$ \tcp*[r]{record selected donor subtrees}

\ForEach{$\mathsf{prog} \in P$}{
    $\widehat{\mathsf{prog}} \leftarrow \mathsf{prog}$\;
    \While{preorder traversing nodes $n$ of $\widehat{\mathsf{prog}}$}{
        $\mathsf{choose} \leftarrow$ \Rand$(0,1)$ \\
        \If{$\mathsf{choose} < w(n)$}{\textbf{continue}} 
        $S \leftarrow \{ d ~|~ d \in \mathsf{Index}[\TypeOf{n}] \land w(d) > \mathsf{choose} \}$ \tcp*[r]{filter target donor nodes}
        \If{$S=\emptyset$}{\textbf{Continue}}
        
        \For{$t \leftarrow 1$ \KwTo $T$}{
            $d \leftarrow \textsc{UniformChoice}(S)$ \tcp*[r]{pick donor node} 

            \If{$d \in \mathsf{Used}$}{
                $w(d) \leftarrow$  \Max{$\gamma \cdot w(d)$, $w_{lb}$} 
            }
            $\widehat{\mathsf{prog}}' \leftarrow \Replace(\widehat{\mathsf{prog}}, n, d)$ \\
            \If{\Check{$\widehat{\mathsf{prog}}'$}}{
                $\mathcal{P}' \leftarrow \mathcal{P}' \cup \{\widehat{\mathsf{prog}}' \} $ \\
                $\mathsf{Used} \leftarrow \mathsf{Used} \cup \{d\}$
            }
            \Else{
                $w(d) \leftarrow w_{lb}$
            }
        }
    }
}

\Return $\mathcal{P}'$

\end{algorithm}

\bocomment{To diversify the seed pool used by the subsequent mutation-based fuzzing stage~\cite{wen2023evaluating,wu2023sjfuzz,gao2024selecting}, we design an adaptive subtree-level mutation algorithm (i.e., one that operates on entire AST subtrees rather than individual nodes or tokens) that adjusts the selection weights of donor subtrees according to whether previous replacements pass language-specific validation.
Its purpose is not to explicitly optimize path novelty or coverage gain, but to efficiently construct more valid and structurally diverse seed programs that can be further mutated by \ourtool.}
Unlike conventional token- or statement-level mutation strategies~\cite{wen2023evaluating,wu2023sjfuzz}, our subtree-level design \bocomment{reduces invalid replacements by restricting each donor subtree to have the same compatibility key as the target subtree}.
\bocomment{Due to the different capabilities and constraints of the underlying AST manipulation libraries, we adopt different validation criteria for the two languages: for Rust, the \texttt{syn} library only accepts programs that can be successfully parsed, and thus we retain mutated programs that pass parsing; for C++, Clang LibTooling operates on compilable programs, and therefore we retain mutated programs that can be successfully compiled. This language-specific validation helps maintain syntactic validity for Rust and compilation feasibility for C++, while encouraging potentially more semantically diverse variants than token- or statement-level mutation.}
\bocomment{Therefore, this component serves only as an auxiliary preprocessing step for enriching the mutation inputs of \ourtool.}
Algorithm~\ref{alg:type-guided-mutation} outlines this procedure.

Given an initial seed set $\mathcal{P}$, the algorithm first builds an AST node type–indexed subtree pool, mapping each AST node type to all subtrees of that type found in $\mathcal{P}$, and assigns each node an initial mutation weight (Lines 1–6). This index serves as a repository of mutation candidates for type-preserving replacements.
The weight of each node is determined by its subtree size, ensuring that smaller subtrees are more likely to be selected while bounding the minimum weight with a user-defined lower limit $w_{lb}$ (Line 6).

For each seed program (Lines 8–26), the algorithm traverses its AST in preorder (Lines 10–26). At each node $n$, it probabilistically decides whether to attempt a mutation based on its weight $w(n)$ (Lines 11–13). If selected, the algorithm retrieves all subtrees of the same AST node type as $n$ from the index whose weights exceed the randomly sampled threshold (Line 14). If no suitable candidates exist, the current node $n$ is skipped (Line 16). Otherwise, the algorithm attempts up to $T$ replacements (Line 17). In each attempt, it randomly samples a donor subtree $d$ from the candidate set (Line 18). If $d$ has been used previously, its weight is decayed (Line 20). The node $n$ is then replaced with $d$, and the mutated program is \bocomment{checked using the corresponding language-specific validation criterion} (Line 22).

\bocomment{If the mutated program passes the corresponding validation criterion, it is added to the output set $\mathcal{P}'$, and $d$ is marked as used (Lines 23--24).}
Otherwise, the weight of $d$ is reset to the lower bound (Line 26). To account for the possibility that $d$ may still be valid in other code contexts where $n$ also appears, we maintain a nonzero lower bound on all weights. Finally, the enhanced seed set $\mathcal{P}'$ is returned, providing a richer and more structurally diverse corpus for the subsequent fuzzing campaign introduced in Section~\ref{sec:main-fuzzing}.

\bocomment{Intuitively, smaller subtrees are more portable across contexts and more likely to remain compilable after replacement, whereas larger subtrees tend to carry stronger structural dependencies. We therefore initialize subtree weights inversely proportional to subtree size.}
The algorithm enforces type compatibility between donor and target subtrees based on their AST node kinds and inferred types.
\bocomment{The algorithm first builds a type-indexed subtree pool by traversing the seed corpus once, which is linear in the total number of AST nodes. It then traverses the ASTs again and attempts replacements only for probabilistically selected nodes. For each selected node, the number of trials is bounded by the retry limit $T$. Thus, the worst-case complexity is $O(|N| + |N| \cdot T \cdot C)$, where $|N|$ is the total number of AST nodes, and $C$ is the validation cost.}

\subsubsection{Examples}

\begin{codelist}[t]
\centering
\begin{minipage}[t]{0.48\linewidth}
\begin{lstlisting}[style=mystyleRust,caption={Rust seed enhancement example.}, captionpos=b, label=lst:rustSeedEnhance]
/* Original */
pub fn foo() {
    let y = 0;
    let x = async || {
        let y = y;
    };
}
/* Enhanced */
pub fn foo() {
    let y = {
        let x = y;
    };
    let x = async || {
        let x = y;
    };
}
\end{lstlisting}
\end{minipage}
\hfill
\begin{minipage}[t]{0.48\linewidth}
\begin{lstlisting}[style=mystyleCpp,caption={C++ seed enhancement example.}, captionpos=b, label=lst:cppSeedEnhance]
/* Original */
static inline unsigned short
swap (unsigned short x) {
  return (x >> 8) 
         | static_cast<unsigned short>(x << 8);
}
/* Enhanced */
static inline unsigned short
swap (unsigned short x) {
  return ((x >> 8) 
         | static_cast<unsigned short>(x << 8) >> 8) 
         | static_cast<unsigned short>(x << 8);
}
\end{lstlisting}
\end{minipage}
\end{codelist}

\bocomment{Listings~\ref{lst:rustSeedEnhance} and~\ref{lst:cppSeedEnhance} illustrate adaptive seed enhancement for Rust and C++, respectively.
Due to the different capabilities of the underlying AST manipulation libraries, the two examples are subject to different validation criteria.
For Rust, the enhanced program is accepted by the \code{syn} parser and thus can be used as a syntactically valid seed for subsequent AST-level mutation, although it may not be successfully compiled by \code{rustc}.
The Rust example introduces a block expression as the initializer of \code{y}.
For C++, Clang LibTooling operates on compilable translation units, and therefore the enhanced program in Listing~\ref{lst:cppSeedEnhance} is retained only after successful compilation.
The C++ example enriches the original bit-manipulation expression with an additional nested shift expression while preserving compilability.
These examples show how seed enhancement expands the seed pool with structurally diverse programs under language-specific validation, i.e., parsing success for Rust and compilation success for C++.}

%% file: Eval.tex
\section{Evaluation}
\label{sec:evaluation}

Our evaluation aims to answer the following research questions:
\begin{enumerate}
    \item[\textbf{RQ1}] How does \ourtool compare to the state-of-the-art compiler fuzzers?

    \item[\bocomment{\textbf{RQ2}}] \bocomment{How does the number of analyzed bug reports affect the effectiveness of \ourtool?}

    \item[\textbf{RQ3}] Is \ourtool effective in finding real-world bugs in production-level compilers?

    \item[\bocomment{\textbf{RQ4}}] \bocomment{How do the key design choices in \ourtool affect its performance?}
    
    \item[\textbf{RQ5}] What are the costs of using \ourtool for mutator synthesis?
    
    \item[\textbf{RQ6}] \bocomment{What are the characteristics and effectiveness of the mutators synthesized by \ourtool?}
\end{enumerate}

\subsection{Implementation} \label{subsec:impl}
To demonstrate the cross-language capability of \ourtool, we select two prominent system-level languages, Rust and C++, as a proof of concept. 
\bocomment{We implement a dedicated web crawler to extract \textit{fixed or resolved} bug reports from the official GitHub repositories of rustc\footnote{\url{https://github.com/rust-lang/rust}} and Clang\footnote{\url{https://github.com/llvm/llvm-project}}, and parse their HTML contents to extract structured fields such as the title of the bug report, bug-triggering program, and compiler outputs/stack traces. The LLM agents operate on these extracted fields rather than on raw HTML pages.}
Both repositories are from production compilers and well-maintained open-source communities.

For each target language, we construct the \textit{initial seed corpus} by combining all extracted bug-triggering programs with the official test suites of the corresponding compilers~\cite{zhong2022enriching}. This process yields 20,481 seeds for Rust and 27,786 seeds for C++.
We expand the corpus by our adaptive seed enhancement approach (i.e., Algorithm~\ref{alg:type-guided-mutation}) for Rust and C++.
In our implementation, we set $\gamma=0.95$, $w_{lb}=0.3$, and $T=200$.
The decay factor $\gamma$ ensures that a node’s selection weight decays to the lower bound after approximately 30 unsuccessful attempts, while the non-zero lower bound $w_{lb}=0.3$ guarantees that every node still retains a small probability of being selected.
Finally, we collect the expanded seed programs, which consist of 87,688 Rust and 68,176 C++ programs, respectively.
Additionally, from the crawled bug reports, we randomly select 500 reports per language for the \textit{Mutator Invention} agent to synthesize language-specific mutators. In total, \ourtool generates 319 valid mutators for Rust and 403 valid mutators for C++.

We employ GPT-4o~\cite{achiam2023gpt} as the underlying LLM, which also offers an API for custom fine-tuning. We utilize this capability to further improve the quality and correctness of the LLM-synthesized mutators.
For fine-tuning, we curated a set of 20 human-written mutators (10 each for Rust and C++) to provide the LLM with high-quality examples of AST-level transformations and \bocomment{more accurate} API usage.
On average, each Rust mutator contains 65 LOC and invokes 16 AST manipulation APIs, while each C++ mutator has 86 LOC and 26 API calls. 
These mutators provide the LLM with a diverse range of high-quality, language-specific demonstrations for learning accurate AST transformation patterns.

We adopted the default supervised fine-tuning (i.e., SFT) pipeline provided by OpenAI\footnote{\url{https://openai.com/index/gpt-4o-fine-tuning/}}. 
\bocomment{The total economic cost of fine-tuning with 20 mutators was 8.62 USD. In our additional experiment with 40 mutators, the fine-tuning cost was 10.72 USD.}
\bocomment{We report this as a separate one-time setup cost and do not include it in the per-mutator synthesis cost reported later.}
Moreover, we configure different temperature values for the agents to balance creativity and determinism: the \textit{Mutator Invention Agent} uses a high temperature (0.8) to encourage diverse and exploratory mutation ideas, while the \textit{Mutator Implementation Synthesis Agent} and \textit{Mutator Refinement Agent} adopt lower temperatures (0.4 and 0.2, respectively) to ensure more stable and consistent code generation.

For the bug-hunting experiments, we initiate the fuzzing loop using all valid mutators.
Within this time budget, we observe that the fuzzing pipelines for both Rust and C++ typically remain in the stage of applying up to two mutators per seed.

\subsection{Evaluated Languages and Compilers}

\begin{table}[tp]
  \centering
    \footnotesize
    \caption{Evaluated languages, compilers, and baseline tools.}
\begin{tabular}{|c|c|l|l|l|l|}
    \hline
    \textbf{Language} & \textbf{Compiler} & \textbf{Fuzzer} & \textbf{Venue} & \textbf{\bocomment{Paradigm}} & \textbf{\bocomment{Core Technique}} \\
    \hline
    \multirow{3}{*}{Rust} & \multirow{3}{*}{rustc/gccrs} & RustSmith~\cite{sharma2023rustsmith}  & ISSTA-23 & \bocomment{Generation-based} & \bocomment{Rust grammar} \\
\cline{3-6}       &    & Rustlantis~\cite{wang2024rustlantis}  & OOPSLA-24 & \bocomment{Generation-based} & \bocomment{Construction rules} \\
\cline{3-6}       &    & Clozemaster~\cite{gao2025clozemaster}  & ICSE-25 & \bocomment{LLM-based} & \bocomment{Masked code completion} \\
    \hline
    \multirow{6}{*}{C++} & \multirow{6}{*}{GCC/Clang} & CSmith~\cite{yang2011finding}  & PLDI-11 & \bocomment{Generation-based} & \bocomment{C/C++ grammar and rules} \\
\cline{3-6}       &    & YARPGen-2.0~\cite{livinskii2023fuzzing} & PLDI-23 & \bocomment{Generation-based} & \bocomment{C/C++ grammar and rules} \\
\cline{3-6}       &    & GrayC~\cite{even2023grayc}  & ISSTA-23 & \bocomment{Mutation-based} & \bocomment{Manually designed mutators} \\
\cline{3-6}       &    & Fuzz4All~\cite{xia2024fuzz4all}  & ICSE-24 & \bocomment{LLM-based} & \bocomment{Direct program generation} \\
\cline{3-6}       &    & MetaMut~\cite{ou2024mutators} & ASPLOS-24 & \bocomment{LLM-based} & \bocomment{LLM-synthesized mutators} \\
\cline{3-6}       &    & TyMut~\cite{wang2025fuzzing} & OOPSLA-25 & \bocomment{Mutation-based} & \bocomment{Manually designed mutators} \\
    \hline
    \end{tabular}
  \tablabel{baselines}
\end{table}

For Rust, we adopt rustc\footnote{\url{https://www.rust-lang.org/}} (the official Rust compiler) and gccrs\footnote{\url{https://github.com/Rust-GCC/gccrs/}} (the GNU front-end for Rust). For C++, we use GCC\footnote{\url{https://gcc.gnu.org/}} and Clang\footnote{\url{https://clang.org/}}, which are the two most widely deployed C++ compilers in practice. 
For the Rust compilers, we adopt the latest release versions, i.e., rustc 1.88 and gccrs-a1a56c6.
For the C++ compilers, we follow the existing study~\cite{ou2024mutators,wang2025fuzzing}, adopting the most recent releases in 2024 (i.e., GCC 14.1.0 and Clang 18.1.0).
\bocomment{To contextualize the line-coverage results, we also report the measured total source-code LOC of each evaluated compiler: 328K LOC for rustc, 97K LOC for gccrs, 1,122K LOC for GCC, and 912K LOC for Clang.}
As test oracles, we apply \textit{crash}, \textit{hang}, and \textit{differential testing} for compilers of the same language~\cite{gao2025clozemaster,zhang2017skeletal}. 
Among them, crash and hang are used to detect compiler failures.
The former indicates a crash or assertion failure, while the latter indicates an infinite loop or deadlock.
Differential testing is used to detect inconsistent behaviors between different compilers, which is a common practice for testing C/C++ compilers that share the same language standard but may have different implementations. For the baseline fuzzers, we use their default configurations from their official repositories. For approaches that require seed programs, we use the same initial seed corpus as \ourtool to ensure a fair comparison.

\subsubsection{Baselines}
\bocomment{To thoroughly evaluate the effectiveness of \ourtool, we select baselines to cover three major compiler-fuzzing paradigms: traditional generation-based fuzzing, mutation-based fuzzing, and recent LLM-based fuzzing. 
The selected tools are publicly available, support the evaluated languages, and represent closely related or state-of-the-art techniques in their respective categories. 
Because most existing compiler fuzzers are language-specific, we choose separate baseline sets for Rust and C++ according to tool availability and language support, as summarized in \tabref{baselines}.}
\bocomment{Unless otherwise specified, all baselines were executed using their publicly available implementations and default or officially recommended configurations.}

\smalltitle{Traditional Fuzzers}
For Rust, we adopt generation-based fuzzers such as RustSmith~\cite{sharma2023rustsmith} and Rustlantis~\cite{wang2024rustlantis}. For C++, we include CSmith~\cite{yang2011finding} and YARPGen~\cite{livinskii2020random,livinskii2023fuzzing}, which follow a similar generation-based paradigm by randomly expanding language grammars and production rules to produce valid programs. We include GrayC~\cite{even2023grayc}, a representative mutation-based C++ fuzzer equipped with manually designed, general-purpose mutators.
Additionally, we adopted the most recent C++ fuzzer, TyMut~\cite{wang2025fuzzing}, which focuses on the type system of C++ compilers.

\smalltitle{LLM-based Fuzzers}
We further compare \ourtool with recent LLM-based compiler fuzzers that follow different design philosophies: we let the LLMs design and implement mutators, which are then applied to the seed programs, but they let the LLMs directly edit the source code to generate mutated programs. For Rust, Clozemaster~\cite{gao2025clozemaster} fine-tunes an LLM to complete masked code segments, enabling direct in-place edits to source programs. For C++, Fuzz4All~\cite{xia2024fuzz4all} utilizes LLMs to rewrite seed programs guided by language standards and official API references. In contrast, MetaMut~\cite{ou2024mutators} uses an LLM to systematically synthesize and combine simple transformation rules into new mutators, falling into the same category as \ourtool but with limited mutator diversity.
\bocomment{We note that the manual repair effort discussed for MetaMut refers to the development process reported in its original paper. In our evaluation, we did not manually repair or modify the outputs of MetaMut or any other baseline; all baselines were run using their publicly available implementations and recommended configurations.}

\subsection{RQ1: Comparison with Existing Fuzzers}
To evaluate \ourtool against existing approaches, we run each fuzzer with a 24-hour time budget, which is a common setting in prior work~\cite{even2023grayc,xia2024fuzz4all,ou2024mutators,yang2024whitefox,wang2025fuzzing}. 
Also, each fuzzer is executed independently for five trials.
\bocomment{For \ourtool, across the five fuzzing runs, the synthesized mutator pool is fixed.}
Following existing studies~\cite{ou2024mutators,wang2025fuzzing}, we report the average line coverage across the five runs, and collect the union of all recovered crashes during all rounds.
For fair comparison, Fuzz4All and MetaMut both use the same backend LLM, GPT-4o, as \ourtool, while Clozemaster is equipped with its original fine-tuned model \bocomment{based on Incoder~\cite{fried2022incoder}}.

\subsubsection{Coverage}

\begin{figure*}[t]
  \centering
    \begin{subfigure}[t]{0.49\linewidth}
    \centering
    \includegraphics[width=\linewidth]{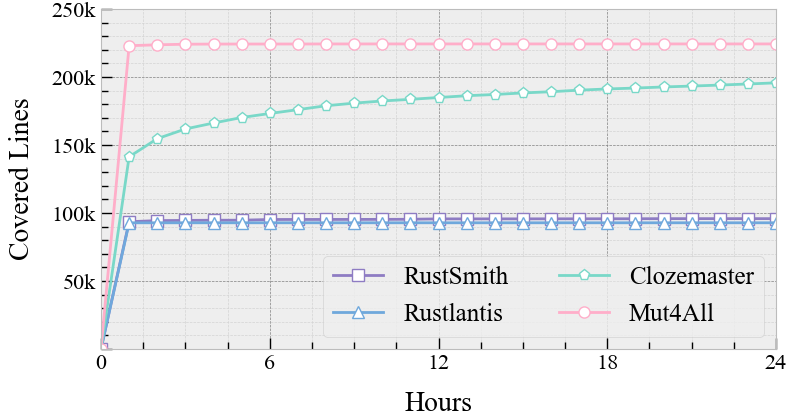}
    \caption{rustc}
    \figlabel{rustc-cov}
  \end{subfigure}
  \begin{subfigure}[t]{0.49\linewidth}
    \centering
    \includegraphics[width=\linewidth]{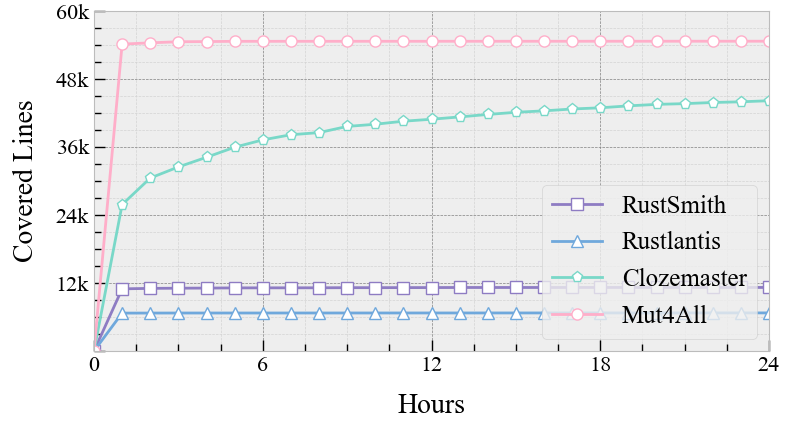}
    \caption{\bocomment{gccrs}}
    \figlabel{gccrs-cov}
  \end{subfigure}
    \caption{\bocomment{The average line coverage of Rust compilers within 5 rounds of 24h-fuzzing.}}
  \figlabel{rust-cov}
\end{figure*}

\begin{figure*}[t]
  \centering
  \begin{subfigure}[t]{0.49\linewidth}
    \centering
    \includegraphics[width=\linewidth]{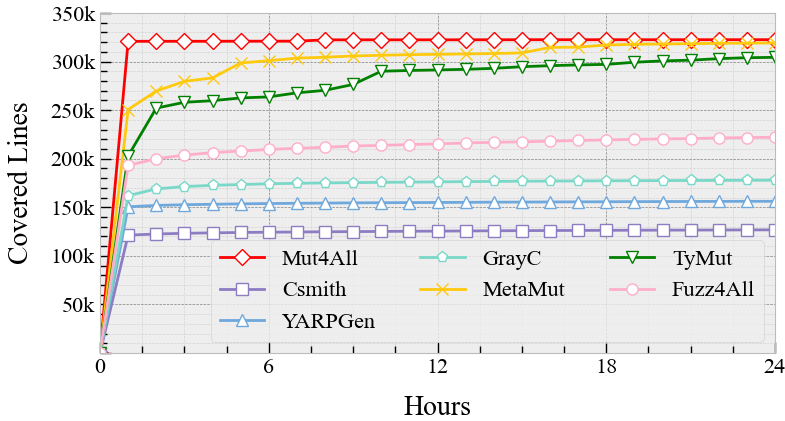}
    \caption{GCC}
    \figlabel{gcc-cov}
  \end{subfigure}
  \begin{subfigure}[t]{0.49\linewidth}
    \centering
    \includegraphics[width=\linewidth]{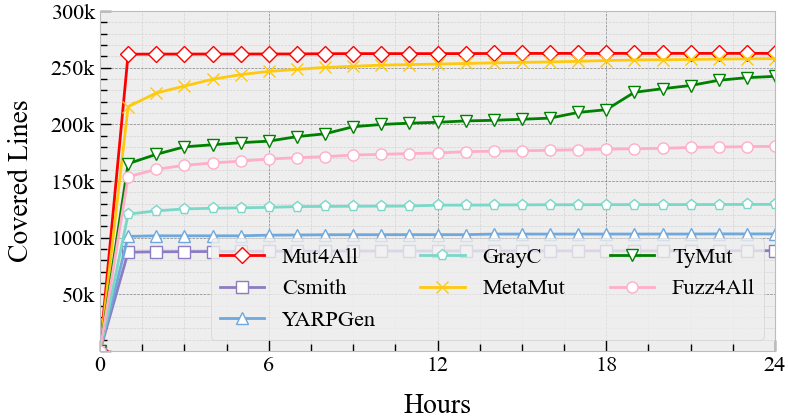}
    \caption{Clang}
    \figlabel{clang-cov}
  \end{subfigure}
    \caption{The average line coverage of C++ compilers within 5 rounds of 24h-fuzzing.}
  \figlabel{c++-cov}
\end{figure*}

\bocomment{We measure compiler source-code line coverage using the coverage tools supported by each compiler’s build infrastructure. For GCC and gccrs, we collect coverage using gcov. For Clang and rustc, we use LLVM instrumentation-based coverage.}

\bocomment{\figref{rust-cov} and \figref{c++-cov} present the average line-coverage trends of all evaluated fuzzers.}
\bocomment{For Rust compilers, \ourtool achieves the highest coverage on both rustc (\figref{rustc-cov}) and gccrs (\figref{gccrs-cov}), consistently} outperforming RustSmith, Clozemaster, and Rustlantis.
For C++ compilers, \ourtool also achieves the highest code coverage on both GCC (\figref{gcc-cov}) and Clang (\figref{clang-cov}), followed by MetaMut and TyMut.
\bocomment{Across all four compilers, \ourtool not only reaches the highest final coverage, but also converges much faster in the early stage of fuzzing, achieving near-saturation substantially earlier than the baselines.
This consistent advantage on both Rust and C++ compilers further demonstrates the effectiveness of \ourtool in exploring diverse compiler behaviors across languages.
Overall, these results show that \ourtool consistently surpasses existing fuzzers in compiler coverage.}

\subsubsection{Unique Crashes}
Following the practice of existing studies~\cite{ou2024mutators,wang2025fuzzing}, a crash is considered unique if its top two stack trace frames—excluding compiler helper functions—differ from those of all other crashes.
We aggregate all unique crashes across the five runs for each fuzzer, as shown in \figref{unique}.
\bocomment{Here, ``exclusive crashes'' refer to unique crashes that are only discovered by \ourtool and not by any baseline.}

\ourtool consistently discovers the highest number of unique crashes for both Rust and C++, significantly surpassing all baselines. For Rust compilers (\figref{rust-unique}), \ourtool uncovers \bocomment{104 crashes that are exclusive to \ourtool, substantially more than the crashes uniquely covered by the baselines.}
For C++ compilers (\figref{cpp-unique}), \bocomment{\ourtool also shows strong complementarity with existing fuzzers, with 13 exclusive crashes.}
\bocomment{The overlaps between \ourtool and each baseline are limited, indicating that \ourtool explores different compiler behaviors from both traditional and LLM-based fuzzers. The results suggest that the mutators synthesized by \ourtool are complementary to existing compiler fuzzing techniques and can expose previously unexplored crash-inducing compiler behaviors.}

\begin{figure}[t]
  \centering
\begin{subfigure}[t]{0.49\linewidth}
    \centering
    \includegraphics[width=\linewidth]{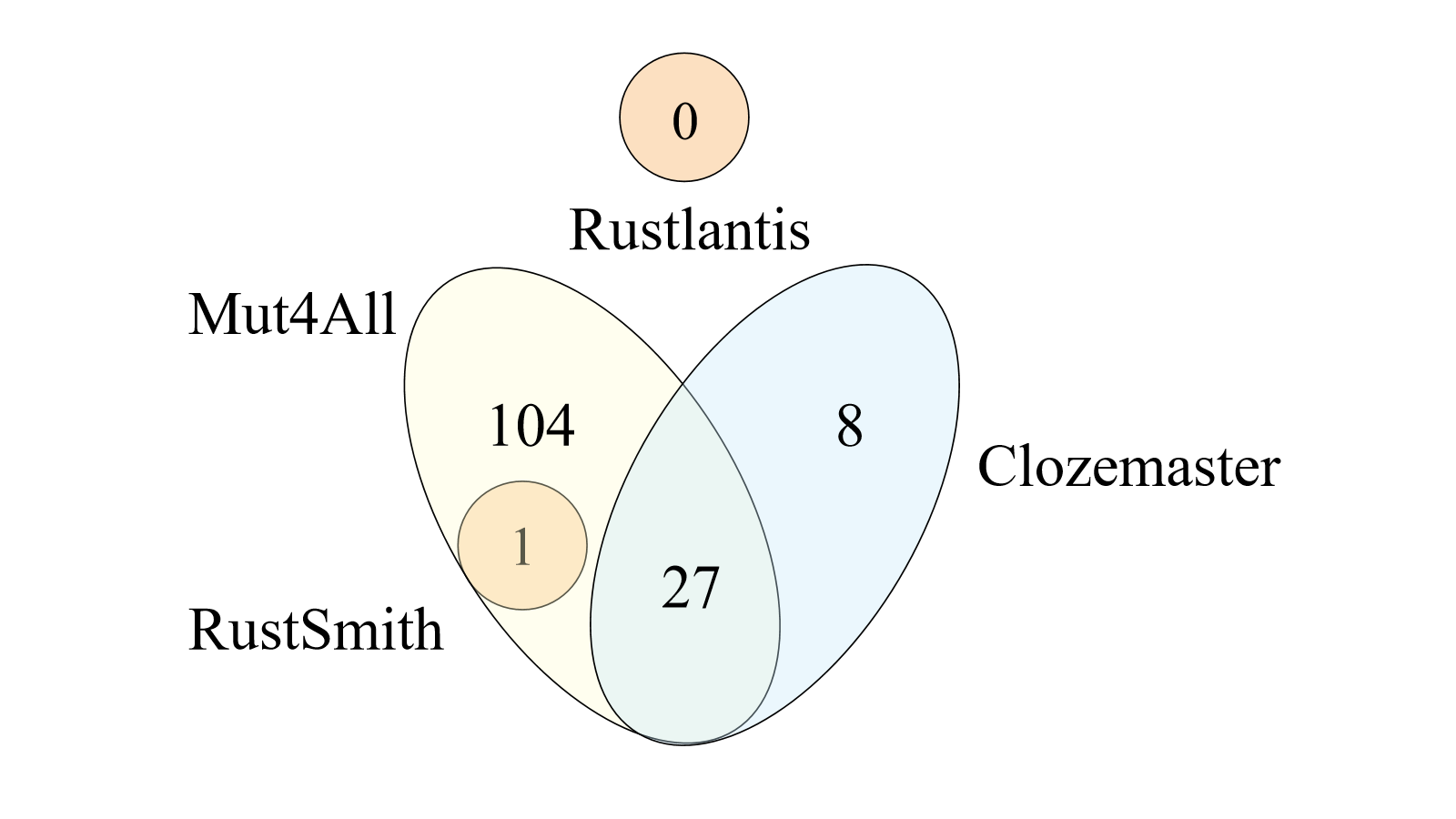}
    \caption{\bocomment{The Rust compilers rustc and gccrs.}}
    \figlabel{rust-unique}
  \end{subfigure}
  \hfill
  \begin{subfigure}[t]{0.49\linewidth}
    \centering
    \includegraphics[width=\linewidth]{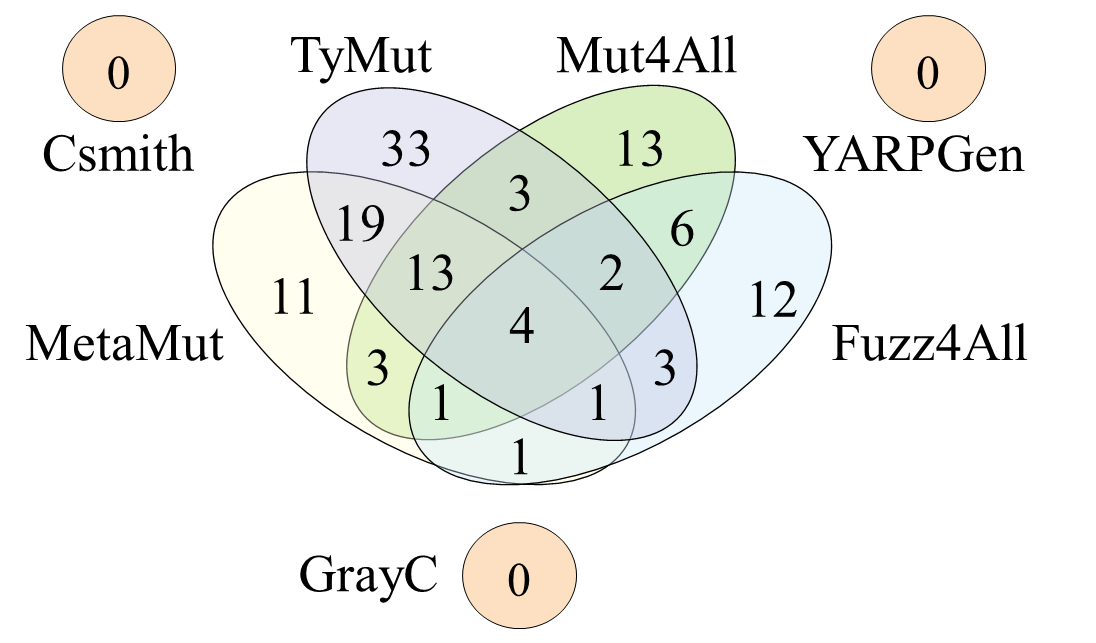}
    \caption{The C++ compilers GCC and Clang.}
    \figlabel{cpp-unique}
  \end{subfigure}
  \caption{Venn diagram of discovered unique crashes.}
  \figlabel{unique}
\end{figure}

\begin{tcolorbox}[boxrule=1pt,left=2pt,right=2pt,top=2pt,bottom=2pt]
\underline{\textbf{Answer to RQ1:}} \ourtool achieves the highest compiler line coverage \bocomment{across all evaluated Rust and C++ compilers}. 
\bocomment{It also discovers many crashes not covered by existing fuzzers, including 104 exclusive crashes in Rust compilers and 13 exclusive crashes in C++ compilers. These results show that \ourtool is highly complementary to existing compiler fuzzers.}
\end{tcolorbox}

\subsection{\bocomment{RQ2: Impact of Bug-Report Budget}}

\bocomment{To evaluate how the number of analyzed bug reports affects the effectiveness of \ourtool, we conduct an incremental bug-report budget analysis.
Specifically, for each language, we divide the originally analyzed 500 bug reports into five cumulative stages with a step size of 100, and further analyze 200 additional reports in this major revision.
These additional reports were sampled from the same source repositories using the same sampling procedure as the original 500 reports.
As a result, for each language, we evaluate seven cumulative settings, ranging from 100 to 700 analyzed bug reports.
For each setting, \ourtool synthesizes mutators based on the corresponding bug-report subset and applies the generated mutators in the subsequent fuzzing campaign.
We measure the impact of the bug-report budget from two perspectives: compiler source-code line coverage and the cumulative number of confirmed new bugs.
Here, the confirmed-bug count is used only as a high-level effectiveness indicator. The detailed status of all submitted bug reports, including confirmed, duplicate, fixed, and unconfirmed cases, is discussed later in RQ3.}

\begin{figure}[t]
  \centering

  \begin{subfigure}[t]{0.49\linewidth}
    \centering
    \includegraphics[width=\linewidth]{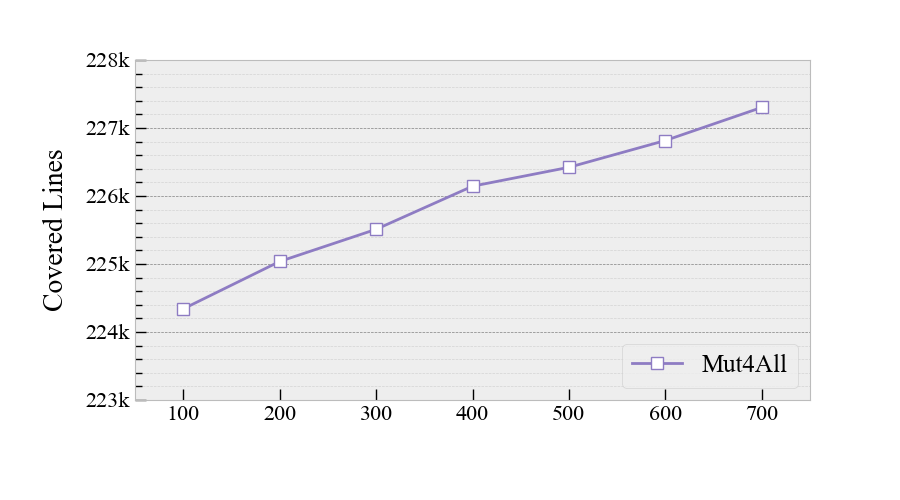}
    \caption{\bocomment{rustc}}
    \figlabel{rustc-report-numbers}
  \end{subfigure}
  \hfill
  \begin{subfigure}[t]{0.49\linewidth}
    \centering
    \includegraphics[width=\linewidth]{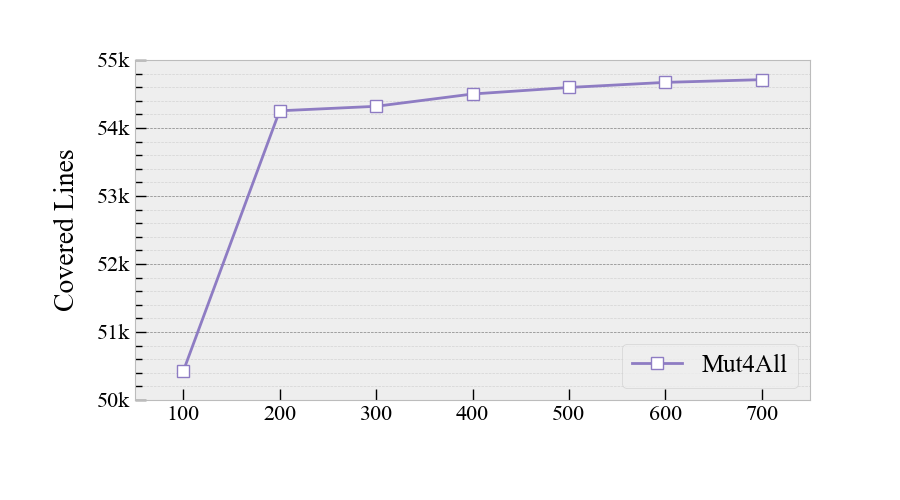}
    \caption{\bocomment{gccrs}}
    \figlabel{gccrs-report-numbers}
  \end{subfigure}

  \begin{subfigure}[t]{0.49\linewidth}
    \centering
    \includegraphics[width=\linewidth]{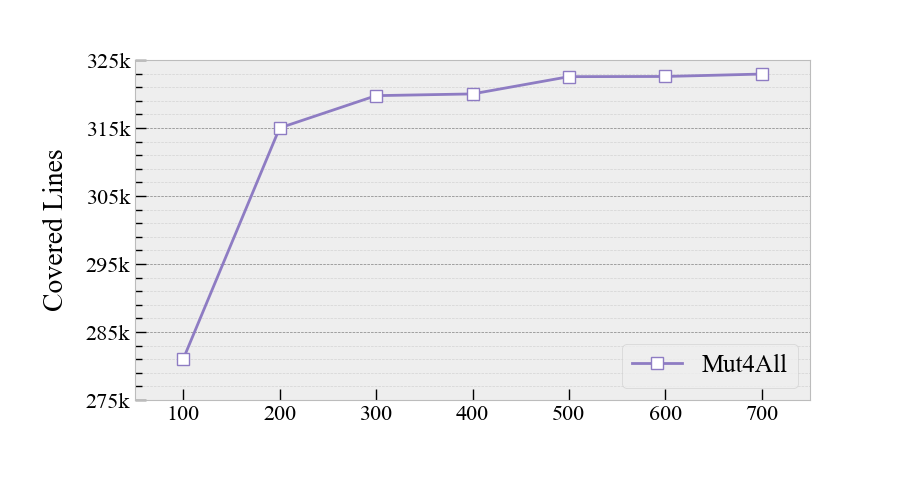}
    \caption{\bocomment{GCC}}
    \figlabel{gcc-report-numbers}
  \end{subfigure}
  \hfill
  \begin{subfigure}[t]{0.49\linewidth}
    \centering
    \includegraphics[width=\linewidth]{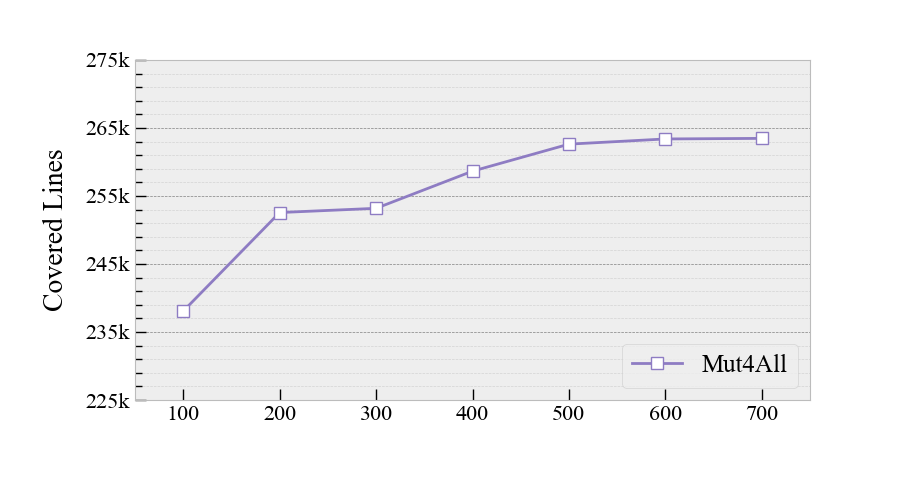}
    \caption{\bocomment{Clang}}
    \figlabel{clang-report-numbers}
  \end{subfigure}
  \caption{\bocomment{Impact of the number of analyzed bug reports on compiler line coverage.}}
  \figlabel{report-num-coverage}
\end{figure}

\begin{table}[tp]
  \centering
  \small
  \caption{\bocomment{Impact of the number of analyzed bug reports on cumulative confirmed new bug discoveries.}}
    \begin{tabular}{|l|r|r|r|r|r|r|r|}
    \hline
       & \bocomment{\textbf{100}} & \bocomment{\textbf{200}} & \bocomment{\textbf{300}} & \bocomment{\textbf{400}} & \bocomment{\textbf{500}} & \bocomment{\textbf{600}} & \bocomment{\textbf{700}} \\
    \hline
    \bocomment{\textbf{Rust Compilers}} & \bocomment{9}  & \bocomment{10} & \bocomment{30} & \bocomment{39} & \bocomment{44} & \bocomment{44} & \bocomment{44} \\
    \hline
    \bocomment{\textbf{C++ Compilers}} & \bocomment{6}  & \bocomment{12} & \bocomment{16} & \bocomment{17} & \bocomment{17} & \bocomment{17} & \bocomment{17} \\
    \hline
    \end{tabular}
  \tablabel{report-num-bug-number}
\end{table}

\bocomment{\figref{report-num-coverage} shows the line coverage achieved under different bug-report budgets. 
Overall, increasing the number of analyzed bug reports consistently improves compiler coverage across all four compilers. 
These results suggest that using more bug reports can broaden the explored compiler code space, even after the initial 500 reports used in the original submission.}

\bocomment{\tabref{report-num-bug-number} presents the cumulative number of confirmed new bugs discovered under different bug-report budgets. 
Different from coverage, the number of confirmed new bugs does not always increase with additional reports. 
For Rust compilers, the number of confirmed new bugs increases from 9 with 100 reports to 44 with 500 reports, but remains unchanged when the budget further increases to 600 and 700 reports. 
Similarly, for C++ compilers, the number increases from 6 to 17 when the budget grows from 100 to 400 reports, but remains unchanged from 400 to 700 reports.}

\bocomment{These results suggest that additional bug reports can improve compiler exploration, as reflected by the steadily increasing line coverage. However, bug discovery is inherently stochastic: adding another batch of reports may broaden the explored compiler behaviors, but does not necessarily expose additional confirmed new bugs. Therefore, the benefit of increasing the bug-report budget is more consistently reflected in coverage improvement than in confirmed bug counts.}

\begin{tcolorbox}[boxrule=1pt,left=2pt,right=2pt,top=2pt,bottom=2pt]
\textbf{Answer to RQ2:} \bocomment{Increasing the number of analyzed bug reports generally improves compiler line coverage, but does not guarantee more confirmed new bugs.}
\end{tcolorbox}

\subsection{RQ3: Bug-Hunting Effectiveness} 

\bocomment{Using the mutators synthesized from 700 bug reports for each language (i.e., the 500 reports used in the original submission plus 200 additional reports analyzed in RQ2), we conducted a long-term fuzzing campaign for each target compiler, including rustc and gccrs for Rust, and GCC and Clang for C++.
Specifically, from the original 500 reports, \ourtool synthesized 319 and 403 valid mutators for Rust and C++, respectively.
The additional 200 reports further contributed 125 and 158 valid mutators for Rust and C++, respectively.
In total, \ourtool synthesized 444 Rust mutators and 561 C++ mutators, which were then used in the bug-hunting experiments.}

\subsubsection{The Status of Our Reported Bugs}
Based on the bug-hunting experiment, we reported a total of \bocomment{100} bugs, as summarized in \tabref{bug-status}.
Note that a unique crash in the previous RQ does not necessarily indicate a newly discovered bug, as some crashes may correspond to the bug reports that have already been fixed in recent compiler versions.
\bocomment{For all abnormal behaviors triggered during fuzzing, including crashes, hangs, and differential behaviors, we manually inspected the generated programs and carefully checked existing bug reports before submission. Therefore, the bugs counted here are not raw tool outputs, but manually verified bug candidates that we judged worthy of reporting.}

Overall, we identified 62 bugs in Rust compilers (44 in rustc and 18 in gccrs) and \bocomment{38} in C++ compilers (\bocomment{24} in GCC and \bocomment{14} in Clang).
\bocomment{Among these, 61 bugs (i.e., 61\%) have been confirmed by compiler maintainers as previously unknown bugs: 30 in rustc, 14 in gccrs, 7 in GCC, and 10 in Clang.}
Notably, \bocomment{35} of the confirmed bugs have already been fixed by the respective compiler developers.
In addition, \bocomment{18} of our reports were classified as duplicates of existing bugs.
\bocomment{These duplicate bug reports were reported previously by other users or developers, rather than being the same bug reports used as input to \ourtool.}
For rustc, 2 bugs involve internal compiler features not intended for public use, and were marked as \textit{Won’t Fix}.
In total, there are \bocomment{19} bugs that remain unconfirmed by the developer across all compilers.
\bocomment{The reported bugs are mainly detected by crash/hang oracles: for Rust, 40 of the 44 reported rustc bugs are crashes and the remaining 4 are hangs; for C++, 14 bugs are crashes and 3 are detected by differential testing.}

These results suggest that \ourtool can uncover a significant number of bugs within a limited time budget, demonstrating the efficiency and effectiveness of our mutator-driven approach.

\begin{table}[tp]
  \centering
  \small
  \captionof{table}{Status of the reported bugs by \ourtool \bocomment{(with 700 bug reports for each language)}.}
    \begin{tabular}{|l|r|r|r|r|r|}
    \hline
    \textbf{Language} & \multicolumn{2}{c|}{\textbf{Rust}} & \multicolumn{2}{c|}{\textbf{C++}} & \multicolumn{1}{c|}{\multirow{2}{*}{\textbf{Total}}} \\
\cline{1-5}    \textbf{Compiler} & \multicolumn{1}{c|}{\textbf{rustc}} & \multicolumn{1}{c|}{\textbf{gccrs}} & \multicolumn{1}{c|}{\textbf{GCC}} & \multicolumn{1}{c|}{\textbf{Clang}} &  \\
    \hline
    \textbf{Reported} & 44 & 18 & \bocomment{24} & \bocomment{14} & \bocomment{100} \\
    \hline
    \textbf{Confirmed as New} & \bocomment{30} & \bocomment{14}  & 7  & 10  & \bocomment{61} \\
    \textbf{Fixed} & \bocomment{18}  & \bocomment{14}  & 2  & 1  & \bocomment{35} \\
    \hline
    \textbf{Duplicate} & \bocomment{12}  & 0  & 2  & \bocomment{4}  & \bocomment{18} \\
    \hline
    \textbf{Won’t Fix} & 2  & 0  & 0  & 0  & 2 \\
    \hline
    \textbf{Unconfirmed} & 0  & \bocomment{4} & \bocomment{15} & 0  & \bocomment{19} \\
    \hline
    \end{tabular}
    \tablabel{bug-status}
\end{table}

\subsubsection{Bug Examples}

We further present several more bugs to show the effectiveness of \ourtool in uncovering subtle, hard-to-construct compiler issues beyond prior fuzzers.

\begin{codelist}[t]
\begin{lstlisting}[style=mystyleRust,caption={The new bug \textit{Rust-142863} uncovered by Mut4All.}, captionpos=b, label=lst:rust142863]
trait Trait<V, D> {}
struct A;
struct B;
// `Trait<V, D>` -> `Trait<NonExistentTrait, NonExistentTrait>`
impl<V, D> Trait<NonExistentTrait, NonExistentTrait> for A
where 
    A: Trait<D, V>, 
    B: Trait<V, D>, 
{}
// `Trait<V, D>` -> `Trait<NonExistentTrait, NonExistentTrait>`
impl<V, D> Trait<NonExistentTrait, NonExistentTrait> for B where A: Trait<V, D> {}
fn impls_trait<T: Trait<V, D>, V, D>() {}
fn main() { impls_trait::<B, _, _>(); }
\end{lstlisting}
\end{codelist}

Listing~\ref{lst:rust142863} shows a Rust program in which valid type parameters in trait implementations are replaced with a deliberately undefined type (i.e., \code{NonExistTrait}). This seemingly illogical mutation causes rustc to enter an infinite compilation loop due to unresolved recursive bounds. Manually crafting such a case would be highly unintuitive, as developers typically avoid inserting undefined types into generic contexts due to their low likelihood of type-checking. However, through automated mutator design and synthesis, \bocomment{\ourtool identifies this transformation and exposes a trait-resolution failure involving cyclic dependencies introduced by invalid substitutions}.

\begin{codelist}[t]
\begin{lstlisting}[style=mystyleCPP,caption={The new bug \textit{LLVM-147009} uncovered by Mut4All.}, captionpos=b, label=lst:llvm147009]
void copy5CharsInto(char *buf)
  __attribute__((enable_if(
    // The original code: `__builtin_object_size(buf, 1) != -1 &&`
    __builtin_object_size(buf, 1024LL) != -1024LL &&
    // The original code: `__builtin_object_size(buf, 1) > 5,`
    __builtin_object_size(buf, 1024LL) > 1024LL,
  "")));
\end{lstlisting}
\end{codelist}

\begin{codelist}[t]
\begin{lstlisting}[style=mystyleCPP,caption={The new bug \textit{LLVM-147170} uncovered by Mut4All.}, captionpos=b, label=lst:llvm147170]
template<typename T>
// `fooT()` -> `fooT(C auto,)`
constexpr int fooT(C auto,) { return __alignof__(T); }
template <typename T>
struct __attribute__((aligned(fooT<T>()))) S { char dummy; };
S<int> s;
\end{lstlisting}
\end{codelist}

Listing~\ref{lst:llvm147009} presents a C++ program in which integer literals used in built-in size checks are replaced with a large edge-case value (\texttt{1024LL}). This subtle transformation triggers a long-standing crash in Clang that had silently persisted since version 9.0, remaining undetected for nearly six years. Although the modified conditions may appear redundant or even nonsensical from a developer's perspective—since the second argument to the built-in checks is typically expected to be between 0 and 3—this mutation exposes latent bugs in how Clang handles unexpected constant values during attribute evaluation. This example illustrates how \ourtool can move beyond human intuitions and systematically explore unconventional input spaces to uncover deep-seated, previously unreachable compiler defects.

Listing~\ref{lst:llvm147170} shows a C++ program combining advanced features such as \texttt{template} instantiation and \texttt{constexpr} evaluation. A mutator generated by \ourtool modifies the valid function declaration \texttt{fooT()} into one with an invalid parameter list: \texttt{fooT(C auto,)}. While this clearly introduces a syntax error, it unexpectedly causes the Clang compiler to crash rather than emit a diagnostic. This subtle transformation exposed a long-standing bug in Clang’s template and attribute evaluation pipeline that had persisted silently since version 9.0. Such non-obvious mutations—where ill-formed syntax is injected into semantically sensitive constructs—are difficult for traditional mutation-based fuzzers, including MetaMut, to uncover. Moreover, the mutated code appears nonsensical to human developers, who would rarely write deliberately malformed parameter lists in \texttt{constexpr} templates. \ourtool’s ability to systematically explore these unintuitive transformations highlights its strength in revealing deeply buried compiler defects.

\subsubsection{\bocomment{Discussion on Bug Importance}}
\bocomment{Although some bugs found by \ourtool are triggered by rare or unintuitive programs, such bugs are still important for compiler testing. 
\ourtool mainly excels at exposing corner-case and robustness bugs involving advanced language features, but these bugs should not be viewed as merely artificial. 
Production compilers are expected to handle unusual and boundary-case inputs robustly, and a crash, hang, or internal compiler error still indicates a real compiler defect. 
In our evaluation, 61 bugs have been confirmed as previously unknown, and 35 have already been fixed by developers, suggesting that many reports are considered important by maintainers. 
Moreover, seemingly rare compiler bugs may still affect real software ecosystems; for example, in our prior work TyMut~\cite{wang2025fuzzing}, a long-undetected Clang bug was later linked by a developer to a TensorFlow issue.}

\begin{tcolorbox}[boxrule=1pt,left=2pt,right=2pt,top=2pt,bottom=2pt]
\underline{\textbf{Answer to RQ3:}} \ourtool discovered \bocomment{100} bugs across four production compilers from Rust and C++, \bocomment{61} of which have been confirmed as previously unknown, demonstrating its effectiveness in hunting real compiler bugs.
Among these confirmed bugs, \bocomment{35} have already been fixed by the developers.
\end{tcolorbox}

\subsection{\bocomment{Ablation Study and Design Choices}}

\subsubsection{\bocomment{Ablation of Agents}}

\bocomment{To understand the contribution of each agent in \ourtool, we conduct an agent-level ablation study.
Since the \textit{Mutator Implementation Synthesis} agent is indispensable for producing executable mutator implementations, we do not remove it.
Instead, we evaluate two ablated variants: 
(1) \ourtool$_{abl-agent1}$, which removes the \textit{Mutator Invention} agent, and 
(2) \ourtool$_{abl-agent3}$, which removes the \textit{Mutator Refinement} agent.
In the first variant, the framework cannot generate bug-report-guided mutator specifications and thus loses the ability to explicitly derive mutation intent from historical bug reports.
In the second variant, raw mutators generated by the synthesis agent are used without feedback-driven repair.
Note that we do not remove the Mutator Implementation Synthesis agent because it is required to generate executable mutator code.}

\begin{figure}[t]
  \centering

  \begin{subfigure}[t]{0.49\linewidth}
    \centering
    \includegraphics[width=\linewidth]{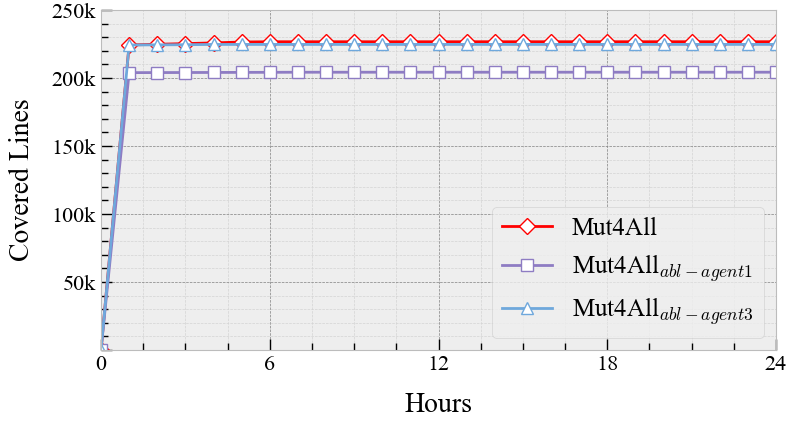}
    \caption{\bocomment{rustc}}
    \figlabel{ablation-agents-rustc}
  \end{subfigure}
  \hfill
  \begin{subfigure}[t]{0.49\linewidth}
    \centering
    \includegraphics[width=\linewidth]{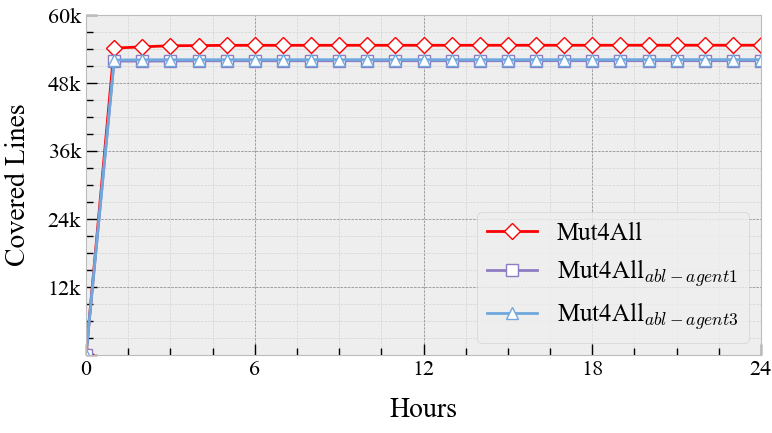}
    \caption{\bocomment{gccrs}}
    \figlabel{ablation-agents-gccrs}
  \end{subfigure}

  \begin{subfigure}[t]{0.49\linewidth}
    \centering
    \includegraphics[width=\linewidth]{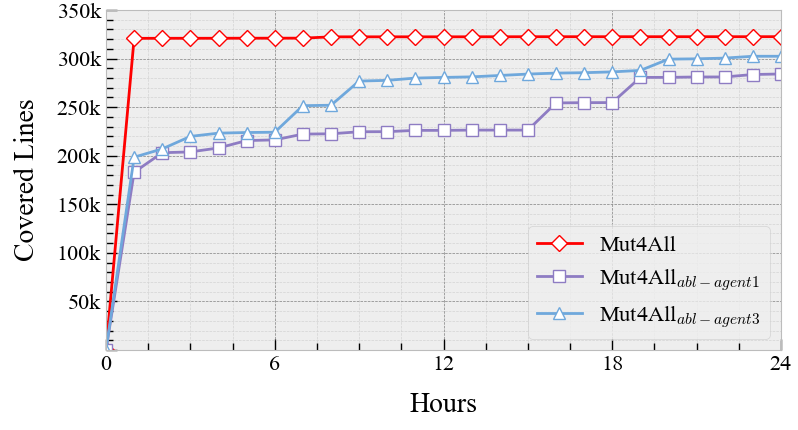}
    \caption{\bocomment{GCC}}
    \figlabel{ablation-agents-gcc}
  \end{subfigure}
  \hfill
  \begin{subfigure}[t]{0.49\linewidth}
    \centering
    \includegraphics[width=\linewidth]{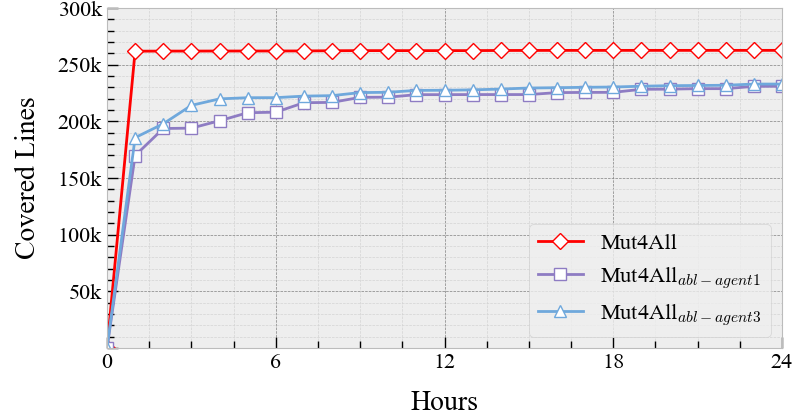}
    \caption{\bocomment{Clang}}
    \figlabel{ablation-agents-clang}
  \end{subfigure}
  
  \caption{\bocomment{Impact of agent ablation on compiler line coverage.}}
  \figlabel{ablation-agents}
\end{figure}

\bocomment{\figref{ablation-agents} shows the line coverage achieved by the full \ourtool and its two ablated variants.
Overall, removing either the \textit{Mutator Invention} agent or the \textit{Mutator Refinement} agent reduces compiler coverage across the evaluated compilers, indicating that both agents contribute to the effectiveness of \ourtool.
The coverage drop is especially clear on GCC and Clang, where the full \ourtool maintains a substantial advantage over both ablated variants throughout the 24-hour fuzzing period.
For rustc and gccrs, the gap is smaller but still observable, suggesting that the two agents also help improve coverage for Rust compilers.

The results also reveal different roles of the two agents.
Removing the \textit{Mutator Invention} agent causes the largest degradation in confirmed-crash reproduction and often leads to larger coverage loss, because the framework loses bug-report-guided mutator specifications and must rely on less targeted mutation designs.
Removing the \textit{Mutator Refinement} agent also hurts performance, since invalid or partially broken mutators cannot be repaired using compiler feedback.
These results confirm that \ourtool benefits from both bug-report-guided mutator invention and feedback-driven mutator refinement.}

\bocomment{In addition to coverage, we also evaluate whether each ablated variant can reproduce the confirmed compiler crashes found by the full \ourtool in RQ3.
Since determining whether two arbitrary bug reports correspond to the same underlying defect can be difficult, we focus this analysis on crashes.
Following the same criterion used for identifying unique crashes in RQ1, we first compare the top two stack frames of each crash, excluding compiler helper functions.
If the stack traces match, we further manually inspect the generated programs and compiler outputs to determine whether they correspond to the same confirmed bug.}

\begin{table}[tp]
  \centering
  \small
  \caption{\bocomment{Effect of ablated \ourtool variants on reproducing confirmed compiler crashes.}}
    \begin{tabular}{|c|c|c|c|c||c||c|}
    \hline
    \multirow{2}{*}{\bocomment{\textbf{Version of \ourtool}}} 
    & \multicolumn{2}{c|}{\bocomment{\textbf{Rust}}} 
    & \multicolumn{2}{c||}{\bocomment{\textbf{C++}}} 
    & \multirow{2}{*}{\bocomment{\textbf{Total}}} 
    & \multirow{2}{*}{\bocomment{\textbf{Decrease vs. Full}}} \\
\cline{2-5}       
    & \bocomment{\textbf{rustc}} 
    & \bocomment{\textbf{gccrs}} 
    & \bocomment{\textbf{GCC}} 
    & \bocomment{\textbf{Clang}} 
    &    &  \\
    \hline
    \bocomment{Full-Featured} 
    & \bocomment{26} & \bocomment{14} & \bocomment{10} & \bocomment{7}  
    & \bocomment{57} & \bocomment{---} \\
    \hline
    \bocomment{No Agent 1} 
    & \bocomment{5}  & \bocomment{10} & \bocomment{1}  & \bocomment{2}  
    & \bocomment{18} & \bocomment{68.4\%} \\
    \bocomment{No Agent 3} 
    & \bocomment{14} & \bocomment{11} & \bocomment{4}  & \bocomment{4}  
    & \bocomment{33} & \bocomment{42.1\%} \\
    \hline
    \bocomment{No Seed Enhance.} 
    & \bocomment{21} & \bocomment{6}  & \bocomment{9}  & \bocomment{6}  
    & \bocomment{42} & \bocomment{26.3\%} \\
    \hline
    \end{tabular}
  \tablabel{confirmed-crashes}
\end{table}

\bocomment{\tabref{confirmed-crashes} shows that the full \ourtool reproduces 57 confirmed compiler crashes in total.
When the \textit{Mutator Invention} agent is removed, the number drops to 18, corresponding to a 68.4\% decrease.
This indicates that bug-report-guided mutator invention is crucial for reaching bug-triggering compiler behaviors.
When the \textit{Mutator Refinement} agent is removed, the number decreases to 33, corresponding to a 42.1\% decrease.
This suggests that feedback-driven repair improves the usability of synthesized mutators and helps reproduce more confirmed crashes.
Overall, these results are consistent with the coverage results: both the invention and refinement agents are important to the bug-finding capability of \ourtool.}

\subsubsection{Ablation of Seed Enhancement}

\bocomment{In this subsection, we study the contribution of the adaptive seed enhancement component.
To this end, we compare the full \ourtool with a variant, denoted as \ourtool$_{NSE}$ (i.e., \textit{N}o \textit{S}eed \textit{E}nhancement), that disables seed enhancement and directly uses the original seed pool for subsequent mutator-based fuzzing.
Note that seed enhancement is not intended to be the main source of bug-finding power in \ourtool.
It serves as an auxiliary preprocessing step to enrich the mutation inputs with more language-specifically validated and structurally diverse programs.}

\begin{figure}[t]
  \centering
  
  \begin{subfigure}[t]{0.49\linewidth}
    \centering
    \includegraphics[width=\linewidth]{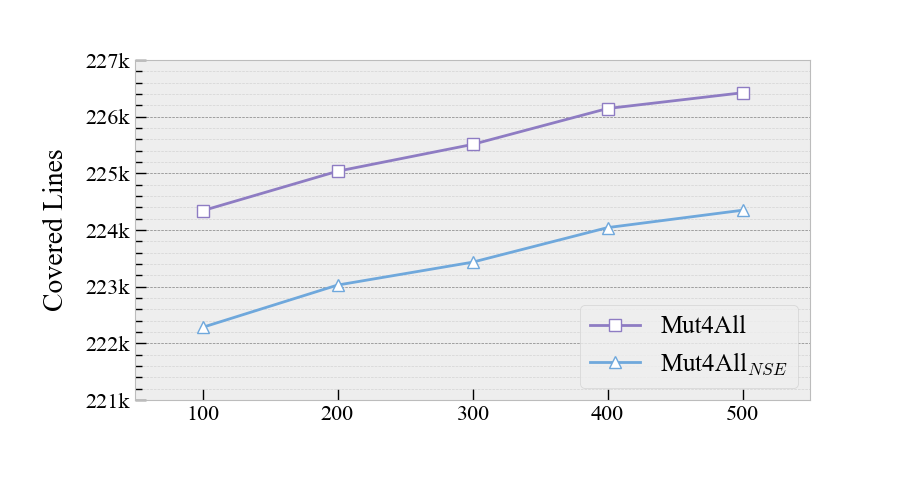}
    \caption{\bocomment{rustc}}
    \figlabel{ablation-seed-rustc}
  \end{subfigure}
  \hfill
  \begin{subfigure}[t]{0.49\linewidth}
    \centering
    \includegraphics[width=\linewidth]{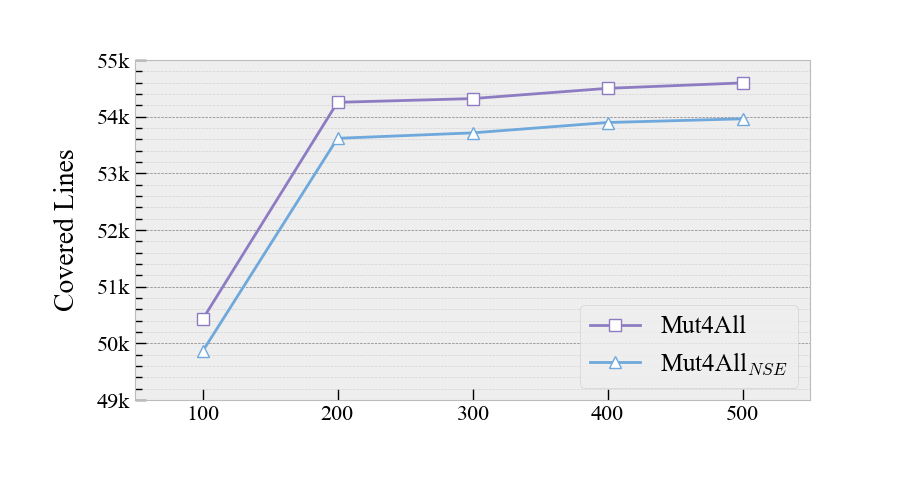}
    \caption{\bocomment{gccrs}}
    \figlabel{ablation-seed-gccrs}
  \end{subfigure}

  \begin{subfigure}[t]{0.49\linewidth}
    \centering
    \includegraphics[width=\linewidth]{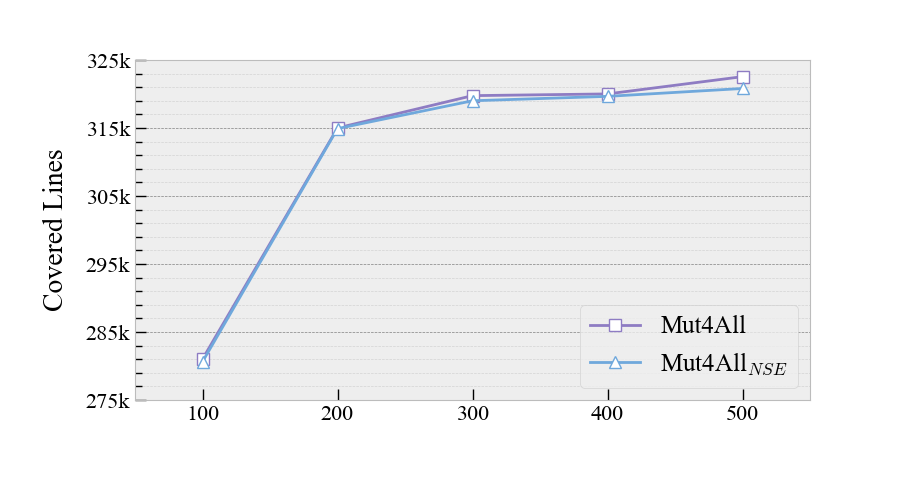}
    \caption{\bocomment{GCC}}
    \figlabel{ablation-seed-gcc}
  \end{subfigure}
  \hfill
  \begin{subfigure}[t]{0.49\linewidth}
    \centering
    \includegraphics[width=\linewidth]{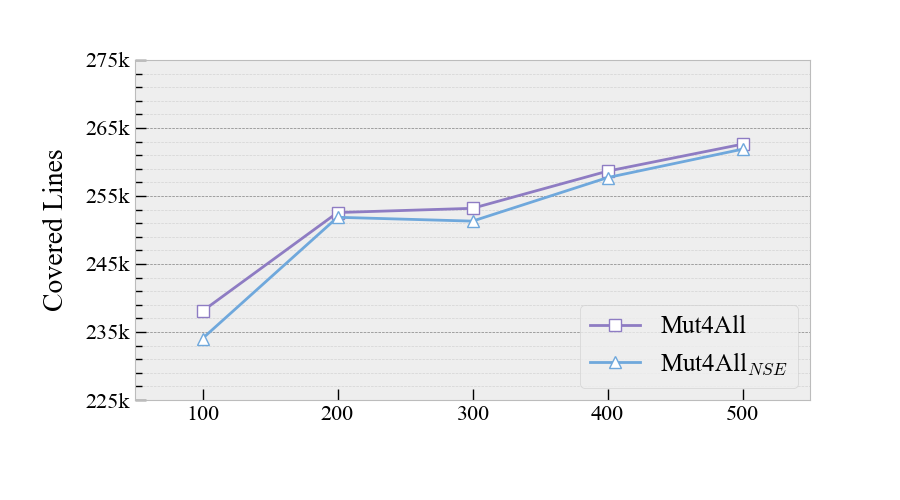}
    \caption{\bocomment{Clang}}
    \figlabel{ablation-seed-clang}
  \end{subfigure}
  
  \caption{\bocomment{Impact of seed enhancement ablation on compiler line coverage.}}
  \figlabel{ablation-seed}
\end{figure}

\bocomment{\figref{ablation-seed} shows the compiler line coverage achieved by \ourtool and \ourtool$_{NSE}$ under different bug-report budgets from 100 to 500.
Overall, enabling seed enhancement consistently improves coverage across all four compilers, although the magnitude of improvement varies.

The results suggest that the adaptive seed enhancement component is beneficial as a lightweight auxiliary step.
By enriching the seed pool with additional structurally diverse programs, it helps the synthesized mutators explore a broader portion of the compiler code base.
At the same time, the relatively modest gaps also confirm that the core bug-finding power of \ourtool still mainly comes from the synthesized mutators, while seed enhancement plays a supportive role.}

\bocomment{We also evaluate the effect of seed enhancement on reproducing confirmed compiler crashes, as reported by the ``No Seed Enhance.'' row in \tabref{confirmed-crashes}.
Without seed enhancement, \ourtool reproduces 42 confirmed crashes, compared with 57 crashes reproduced by the full version, corresponding to a 26.3\% decrease.
This result shows that seed enhancement helps \ourtool reach more bug-triggering compiler behaviors.
However, its impact is smaller than removing either the \textit{Mutator Invention} or \textit{Mutator Refinement} agent, which is consistent with our design: seed enhancement is an auxiliary component for enriching mutation inputs, while the main bug-finding capability comes from the synthesized mutators.}

\subsubsection{\bocomment{Sensitivity to the Configuration of Seed Enhancement}}

\bocomment{We evaluate the sensitivity of \ourtool to the configuration of adaptive seed enhancement.
Recall that the seed enhancement algorithm has three main parameters: the decay factor $\gamma$, the lower bound of subtree weights $w_{lb}$, and the retry limit $T$.
Our default setting is $(\gamma, w_{lb}, T)=(0.95, 0.3, 200)$.
To study the impact of these parameters, we vary one parameter at a time while keeping the other two unchanged.
In total, we evaluate seven configurations:
$(0.95, 0.3, 200)$, $(0.95, 0.1, 200)$, $(0.95, 0.5, 200)$, $(0.95, 0.3, 100)$, $(0.95, 0.3, 300)$, $(0.90, 0.3, 200)$, and $(1.00, 0.3, 200)$.
For each configuration, we generate 50,000 enhanced seed programs and repeat the fuzzing experiment five times.
We use average compiler line coverage and the number of unique crashes as evaluation metrics.}

\begin{figure}[t]
  \centering
    \begin{subfigure}[t]{0.49\linewidth}
    \centering
    \includegraphics[width=\linewidth]{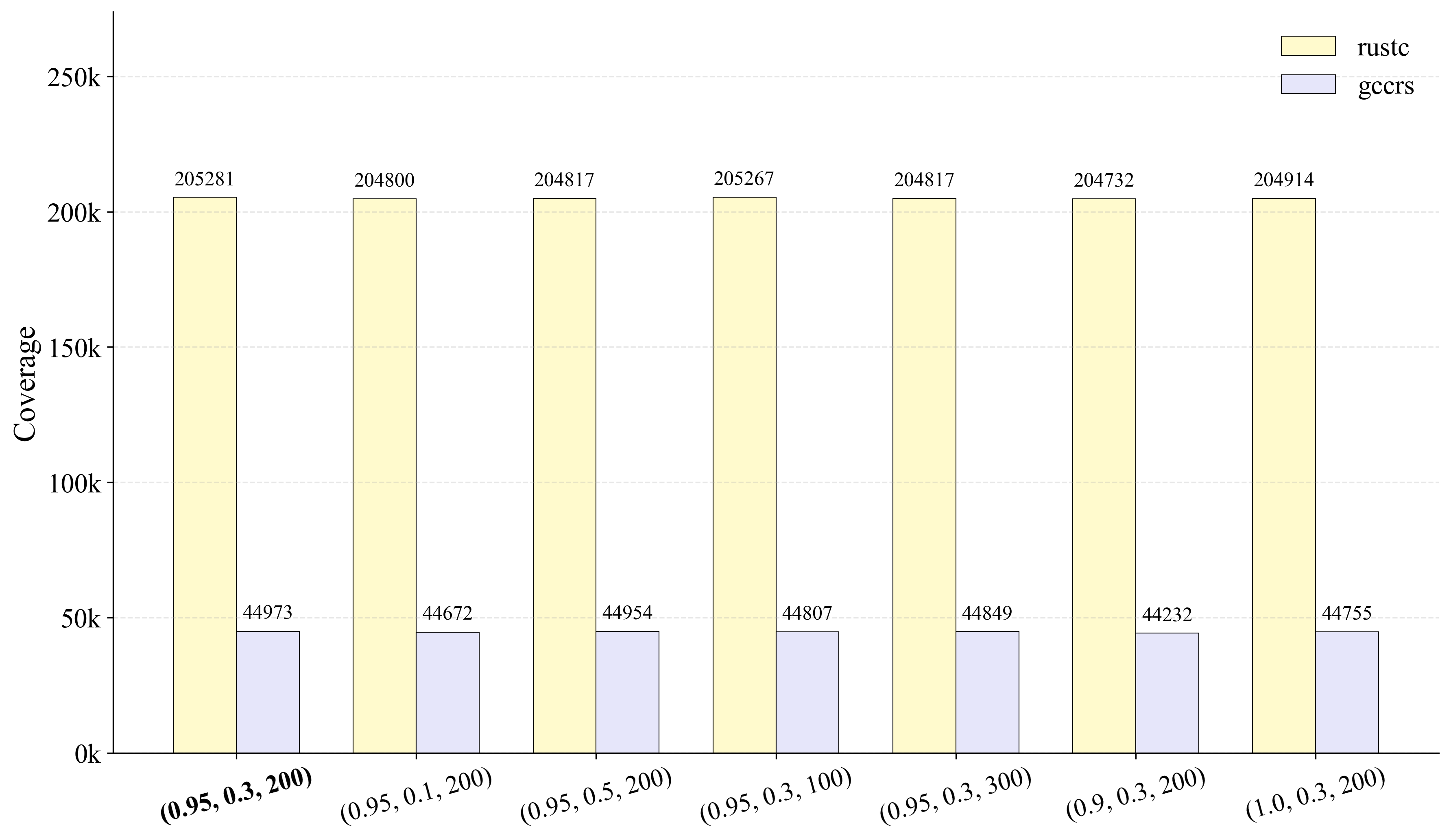}
    \caption{\bocomment{Rust}}
    \figlabel{seed-conf-cov-rust}
  \end{subfigure}
  \hfill
  \begin{subfigure}[t]{0.49\linewidth}
    \centering
    \includegraphics[width=\linewidth]{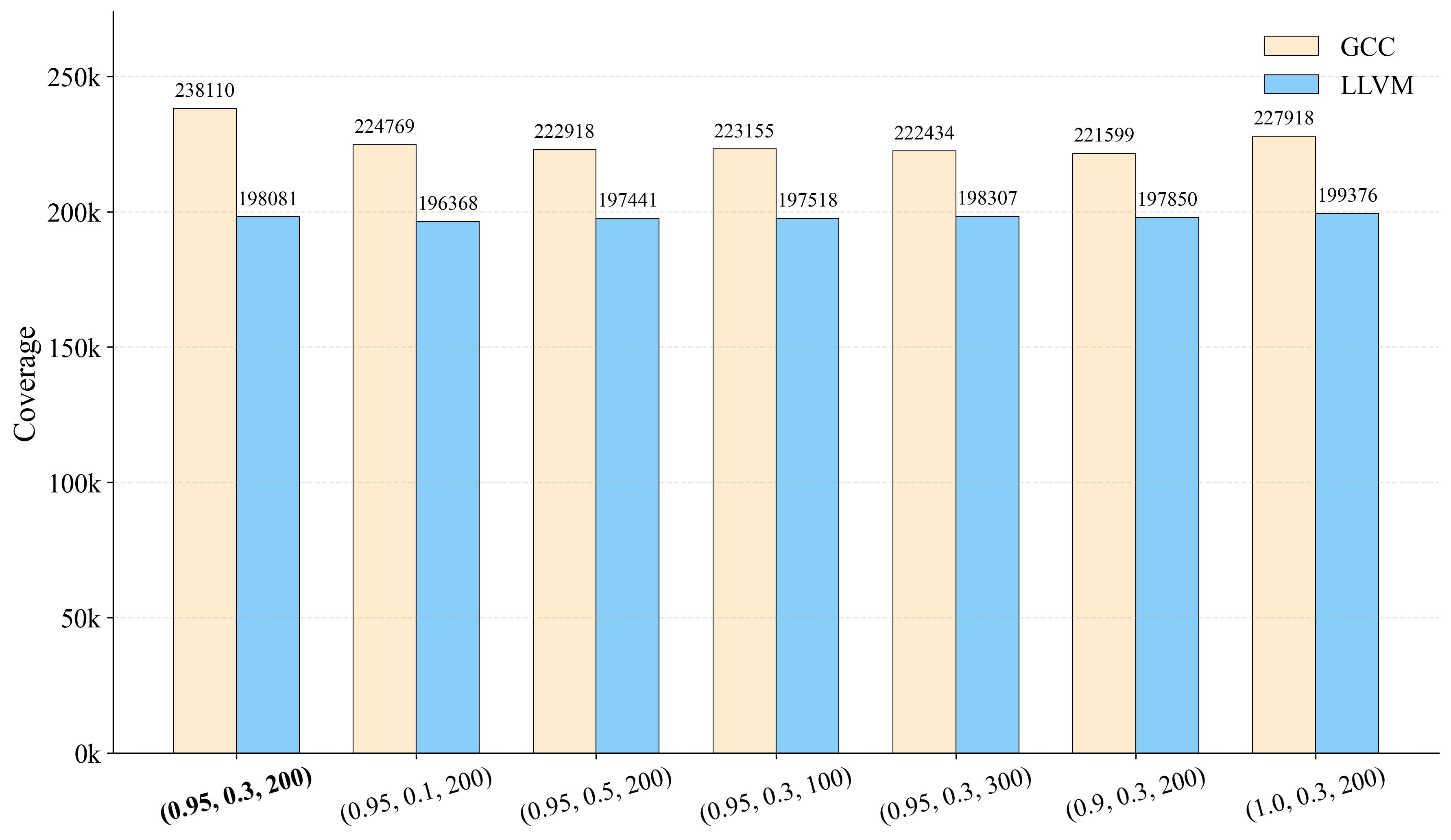}
    \caption{\bocomment{C++}}
    \figlabel{seed-conf-cov-c++}
  \end{subfigure}
  
  \caption{\bocomment{Impact of seed enhancement configurations on compiler line coverage.}}
  \figlabel{seed-conf-cov}
\end{figure}

\begin{figure}[t]
  \centering
  \begin{subfigure}[t]{0.49\linewidth}
    \centering
    \includegraphics[width=\linewidth]{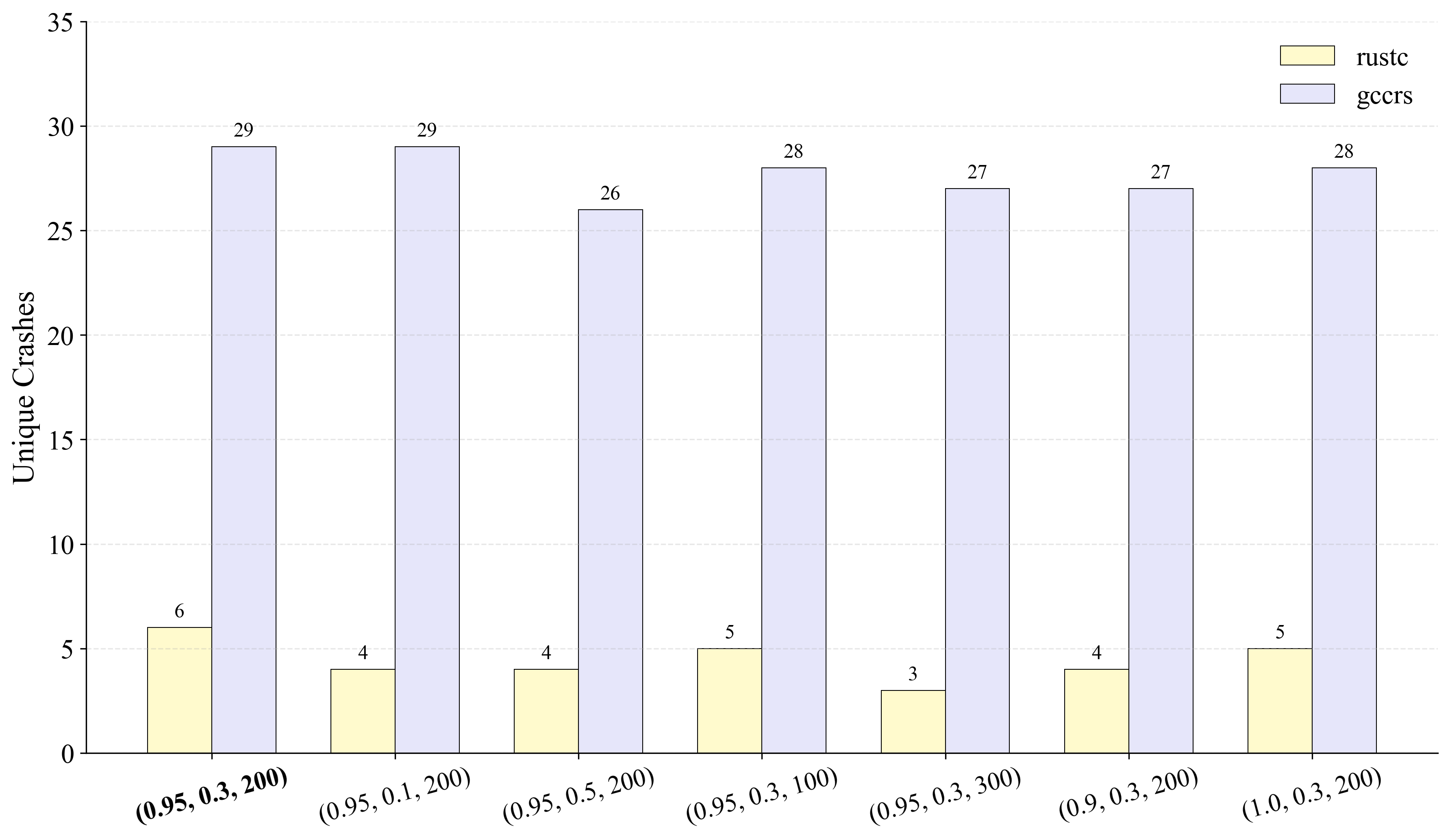}
    \caption{\bocomment{Rust}}
    \figlabel{seed-conf-uc-rust}
  \end{subfigure}
  \hfill
   \begin{subfigure}[t]{0.49\linewidth}
    \centering
    \includegraphics[width=\linewidth]{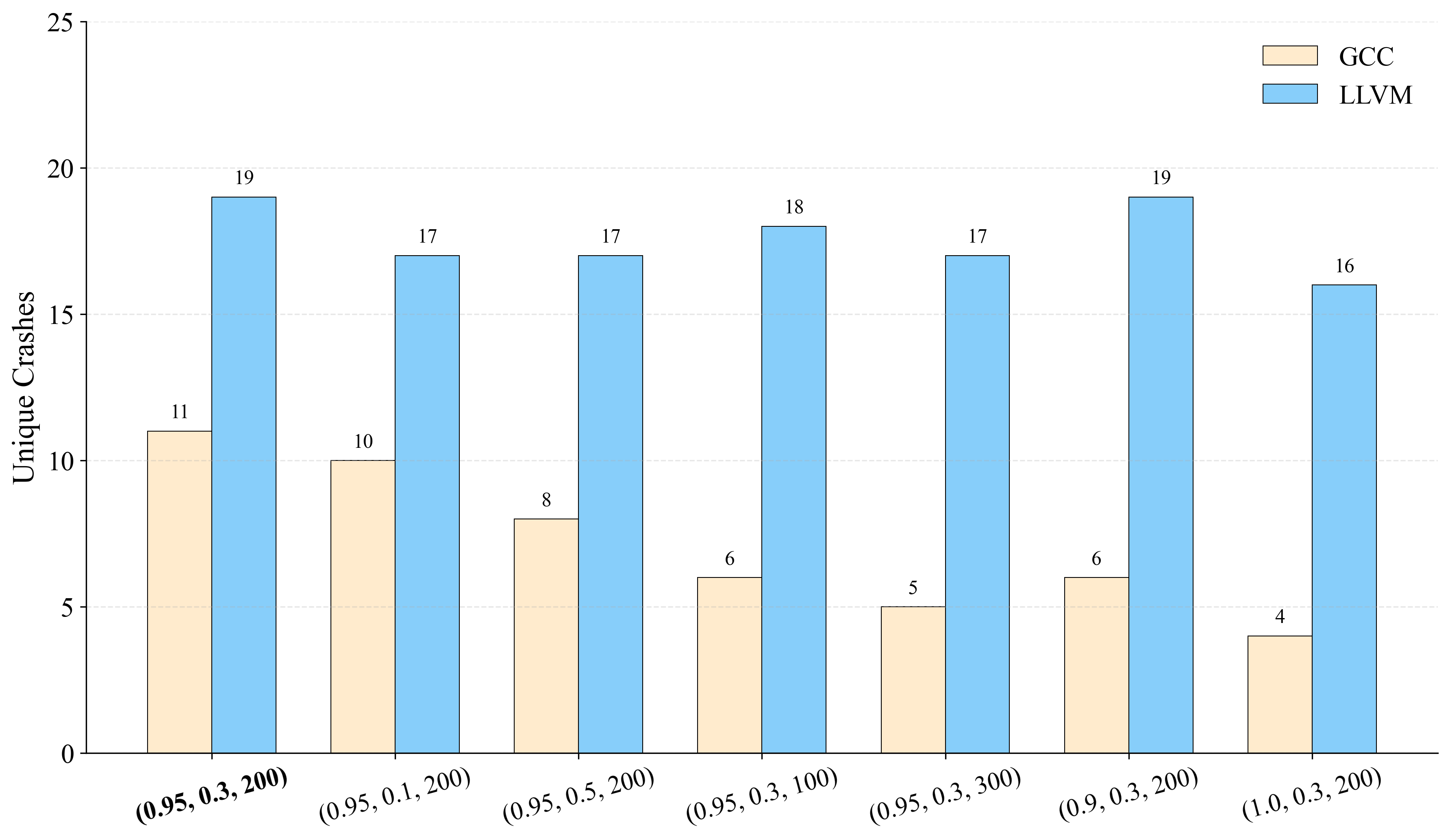}
    \caption{\bocomment{C++}}
    \figlabel{seed-conf-uc-c++}
  \end{subfigure}
  
  \caption{\bocomment{Impact of seed enhancement configurations on unique crashes.}}
  \figlabel{seed-conf-uc}
\end{figure}

\bocomment{\figref{seed-conf-cov} and \figref{seed-conf-uc} show that the default configuration achieves the best or near-best overall performance across the evaluated compilers.
Although some alternative settings slightly improve a specific metric on a given compiler, none consistently outperform the default setting across both languages.
The results also show that coverage and unique crashes vary only moderately across configurations, suggesting that \ourtool is not highly sensitive to the seed enhancement parameters.}

\subsubsection{\bocomment{Sensitivity to LLM-Related Configurations}}

\bocomment{In this subsection, we evaluate how LLM-related configurations influence the mutator synthesis effectiveness of \ourtool. Specifically, we study the following aspects: (1) the effect of LLM temperature, (2) the impact of using different LLMs, and (3) the influence of the number of fine-tuning examples. These experiments aim to assess the robustness of \ourtool to LLM configuration choices and to understand how these choices affect valid-mutator synthesis.}

\smalltitle{\bocomment{Sensitivity to Temperature}}
\bocomment{We study the impact of LLM temperature on mutator synthesis. Recall that \ourtool uses three LLM agents, and their default temperatures are set to $(0.8, 0.4, 0.2)$ for the \textit{Mutator Invention}, \textit{Mutator Implementation Synthesis}, and \textit{Mutator Refinement} agents, respectively.
To evaluate whether this configuration is appropriate, we randomly sample 50 bug reports from the original 500 reports for each language and run \ourtool under different temperature configurations.
We vary the temperature of one agent at a time while keeping the other two unchanged, resulting in 10 configurations in total.
For each configuration, we measure the number of generated mutators, valid mutators, and invalid mutators.}

\begin{figure}[t]
  \centering
    \begin{subfigure}[t]{0.49\linewidth}
    \centering
    \includegraphics[width=\linewidth]{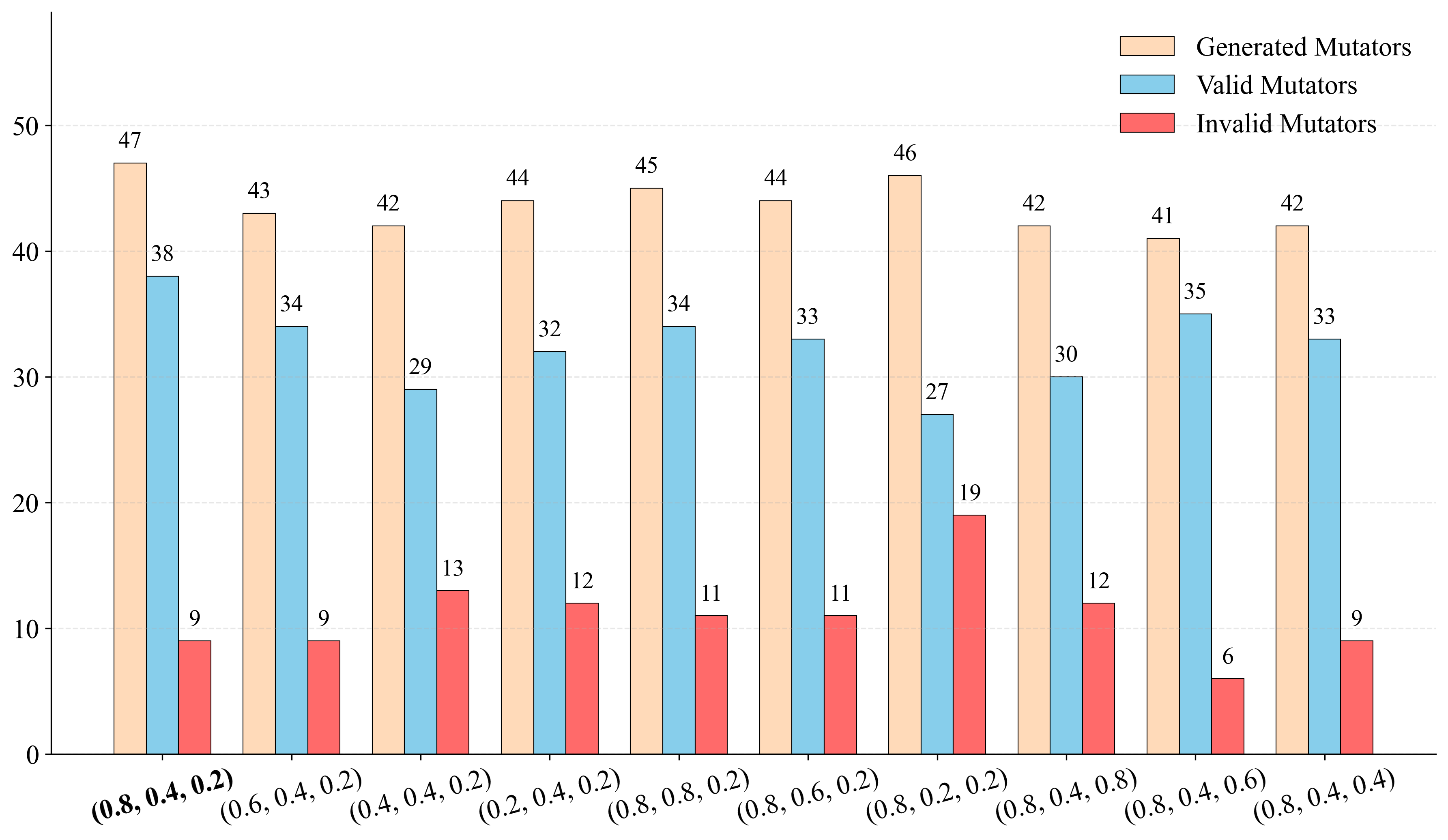}
    \caption{\bocomment{Rust}}
    \figlabel{temperature-rust}
  \end{subfigure}
  \hfill
  \begin{subfigure}[t]{0.49\linewidth}
    \centering
    \includegraphics[width=\linewidth]{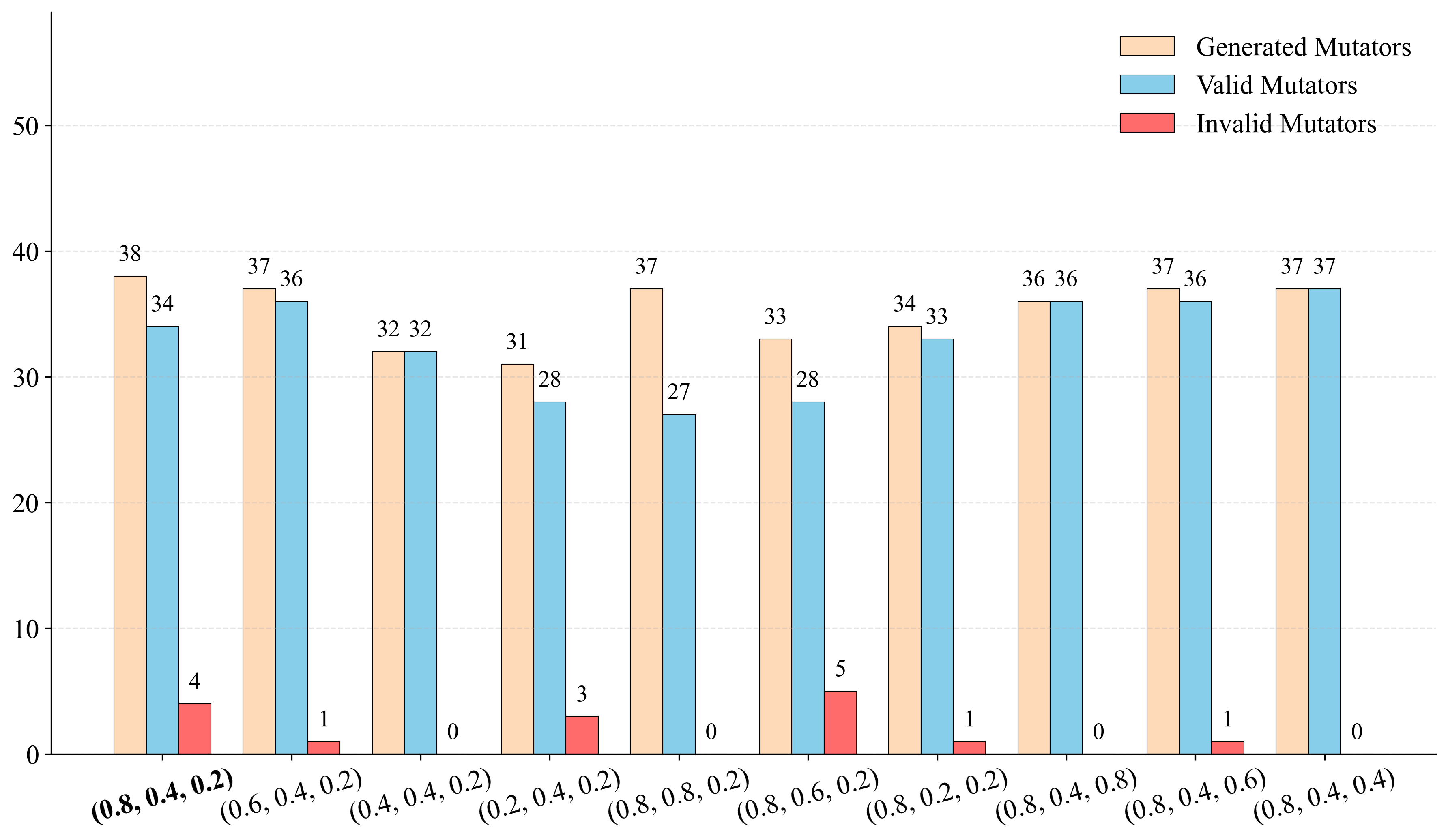}
    \caption{\bocomment{C++}}
    \figlabel{temperature-c++}
  \end{subfigure}
  
  \caption{\bocomment{Effect of LLM temperature across the three agents.}}
  \figlabel{temperature}
\end{figure}

\bocomment{As shown in \figref{temperature}, the default configuration $(0.8, 0.4, 0.2)$ achieves the best overall performance in terms of valid-mutator synthesis in both languages.
It produces the largest number of generated mutators and valid mutators for both Rust and C++.
Changing the temperature of any single agent does not consistently improve the number of valid mutators, and in several cases leads to more invalid mutators.
This indicates that the default setting provides a reasonable balance between exploration and stability: a relatively higher temperature helps the invention agent generate diverse mutator ideas, while lower temperatures help the synthesis and refinement agents produce more stable and correct code.}

\smalltitle{\bocomment{Sensitivity to Different LLMs}}
\bocomment{To evaluate the impact of the underlying LLM on mutator synthesis, we replace the default GPT-4o with three recent open-source models: DeepSeek-V3.2 685B, Llama-3.3 70B Instruct, and Qwen3 32B.
Using the same 500 bug reports for each language, we generate mutators with each model and count the number of valid mutators successfully produced.}

\begin{table}[tp]
  \centering
  \small
  \caption{\bocomment{Valid mutators synthesized by different LLMs.}}
    \begin{tabular}{|l|c|c||c|}
    \hline
     \bocomment{\textbf{LLM}}   & \bocomment{\textbf{Rust}} & \bocomment{\textbf{C++}} & \bocomment{\textbf{Sum}} \\
    \hline
    \bocomment{GPT-4o (Default)} & \bocomment{319} & \bocomment{403} & \bocomment{722} \\
    \hline
    \bocomment{Qwen3 32B} & \bocomment{278} & \bocomment{280} & \bocomment{558} \\
    \bocomment{DeepSeek-V3.2 685B} & \bocomment{235} & \bocomment{297} & \bocomment{532} \\
    \bocomment{Llama-3.3 70B} & \bocomment{186} & \bocomment{264} & \bocomment{450} \\
    \hline
    \end{tabular}
  \tablabel{llm-comparison}
\end{table}

\bocomment{The results are summarized in \tabref{llm-comparison}.
We observe that GPT-4o consistently generates the highest number of valid mutators for both Rust and C++, while the open-source models produce fewer valid outputs.
These results highlight that the choice of LLM significantly affects mutator generation quality and that GPT-4o currently provides the best performance in terms of valid mutator synthesis.}

\bocomment{To further understand whether the valid mutators synthesized by different backend LLMs are also effective in downstream fuzzing, we evaluate their ability to reproduce the confirmed compiler crashes found by the default GPT-4o-based \ourtool.
Following the same criterion used in the ablation study, we focus on crashes and compare the top two stack frames, followed by manual inspection to determine whether they correspond to the same confirmed bug.}

\begin{table}[tp]
  \centering
  \small
  \caption{\bocomment{Effect of different backend LLMs on reproducing confirmed compiler crashes.}}
    \begin{tabular}{|c|c|c|c|c||c||c|}
    \hline
    \multirow{2}{*}{\bocomment{\textbf{Backend Model}}} 
    & \multicolumn{2}{c|}{\bocomment{\textbf{Rust}}} 
    & \multicolumn{2}{c||}{\bocomment{\textbf{C++}}} 
    & \multirow{2}{*}{\bocomment{\textbf{Total}}} 
    & \multirow{2}{*}{\bocomment{\textbf{Decrease vs. GPT-4o}}} \\
\cline{2-5}       
    & \bocomment{\textbf{rustc}} 
    & \bocomment{\textbf{gccrs}} 
    & \bocomment{\textbf{GCC}} 
    & \bocomment{\textbf{Clang}} 
    &    &  \\
    \hline
    \bocomment{GPT-4o (Default)} 
    & \bocomment{26} & \bocomment{14} & \bocomment{10} & \bocomment{7}  
    & \bocomment{57} & \bocomment{---} \\
    \hline
    \bocomment{Qwen3 32B} 
    & \bocomment{11} & \bocomment{11} & \bocomment{0} & \bocomment{2}  
    & \bocomment{24} & \bocomment{57.9\%} \\
    \bocomment{DeepSeek-V3.2 685B} 
    & \bocomment{11} & \bocomment{11} & \bocomment{3} & \bocomment{4}  
    & \bocomment{29} & \bocomment{49.1\%} \\
    \bocomment{Llama-3.3 70B} 
    & \bocomment{9} & \bocomment{10} & \bocomment{0} & \bocomment{0}  
    & \bocomment{19} & \bocomment{66.7\%} \\
    \hline
    \end{tabular}
  \tablabel{confirmed-crashes-llm}
\end{table}

\bocomment{\tabref{confirmed-crashes-llm} shows that GPT-4o also achieves the best downstream bug-reproduction effectiveness.
Compared with GPT-4o, the open-source models reproduce substantially fewer confirmed compiler crashes.
Among them, DeepSeek-V3.2 performs the best, reproducing 29 confirmed crashes, followed by Qwen3 with 24 and Llama-3.3 with 19.
These results are consistent with the valid-mutator synthesis results in \tabref{llm-comparison}, indicating that the backend LLM affects not only the number of valid mutators, but also their downstream fuzzing effectiveness.
Overall, GPT-4o remains the strongest backend model in our current evaluation.}

\smalltitle{\bocomment{Sensitivity to Fine-Tuning}}
\bocomment{To study the impact of fine-tuning on mutator synthesis, we evaluate three configurations for the mutator implementation synthesis agents:
\begin{enumerate}
    \item \textit{No fine-tuning} (0 manually written mutators),
    \item \textit{Default configuration: 10+10} (10 Rust manually written mutators + 10 C++ manually written mutators),
    \item \textit{20+20} (20 Rust manually written mutators + 20 C++ manually written mutators).
\end{enumerate}
Each configuration is evaluated on the same bug reports (i.e., the original 500 for each language), counting the number of valid mutators successfully generated.}

\begin{table}[tp]
  \centering
  \small
  \caption{\bocomment{Effect of fine-tuning on synthesizing valid mutators (default: 10+10).}}
    \begin{tabular}{|l|c|c||c|}
    \hline
       \bocomment{\textbf{Configuration}} & \bocomment{\textbf{Rust}}  & \bocomment{\textbf{C++}} & \bocomment{\textbf{Sum}} \\
    \hline
    \bocomment{No fine-tuning (0)}  & \bocomment{306} & \bocomment{392} & \bocomment{698} \\
    \hline
    \bocomment{Default (10+10)} & \bocomment{319} & \bocomment{403} & \bocomment{722} \\
    \hline
    \bocomment{Large (20+20)} & \bocomment{247} & \bocomment{379} & \bocomment{626} \\
    \hline
    \end{tabular}
  \tablabel{sft-valid-mutators}
\end{table}

\bocomment{From \tabref{sft-valid-mutators}, we observe that the default \textit{10+10} configuration yields the highest number of valid mutators for both Rust and C++, confirming that a moderate number of high-quality fine-tuning examples provides the best balance for guiding LLM synthesis.
Increasing to \textit{20+20} slightly reduces the number of valid mutators. This result suggests that the quality and representativeness of fine-tuning examples may be more important than their quantity.}

\bocomment{These results do not imply that fine-tuning guarantees correct AST API usage or faithful transformations. Rather, they show that a compact set of representative demonstrations improves the likelihood of generating valid mutators, while validation and refinement remain necessary to filter or repair incorrect outputs.}

\subsubsection{\bocomment{Sensitivity to Refinement Rounds}} \label{sec:eval-refinement-rounds}
\bocomment{The limit of the number of refinement rounds is critical of the \textit{Mutator Refinement} agent. Therefore, we evaluate the effect of the number of refinement rounds on the synthesis of valid mutators. Using the original 500 bug reports per language, we vary the maximum refinement rounds from 0 to 20 and measure the number of valid mutators generated.

As shown in \figref{refinement-rounds}, increasing refinement rounds generally produces more valid mutators for both Rust and C++. The gain is most pronounced in the initial rounds and gradually saturates, suggesting diminishing returns beyond approximately 10–15 rounds. This indicates that multiple refinement iterations improve the correctness and usability of synthesized mutators, and most gains occur within the first 10 rounds, while further refinement brings only marginal improvements.}

\begin{figure*}[t!]
    \centering
    \includegraphics[width=0.6\linewidth]{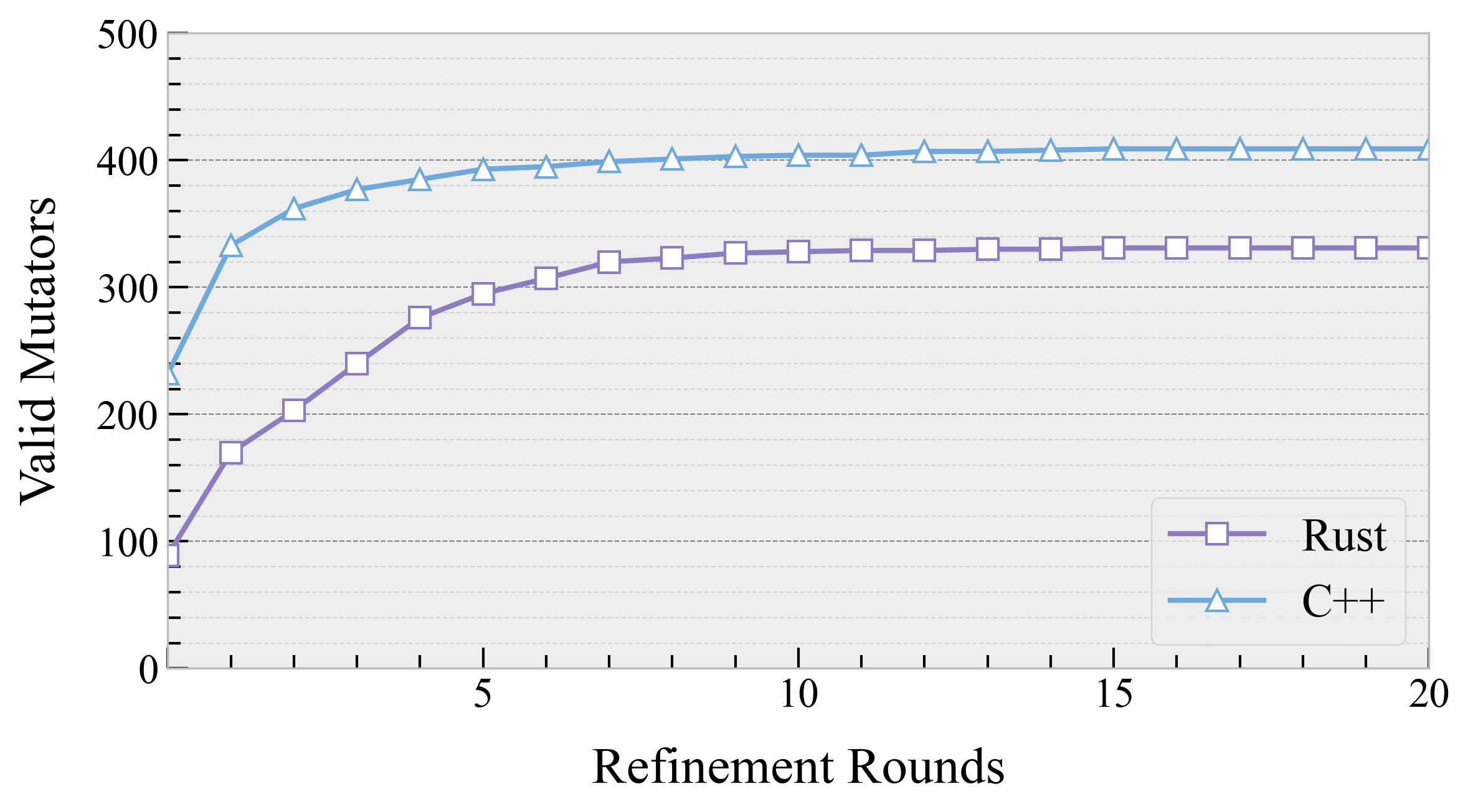}
    \caption{\bocomment{Number of valid mutators generated with different refinement rounds.}}
    \figlabel{refinement-rounds}
\end{figure*}
\begin{tcolorbox}[boxrule=1pt,left=2pt,right=2pt,top=2pt,bottom=2pt]
\underline{\textbf{Answer to RQ4:}} \bocomment{\ourtool benefits from both bug-report-guided mutator invention and feedback-driven refinement, and seed enhancement provides a useful auxiliary improvement in coverage. 
The default seed-enhancement and LLM-related configurations are generally among the best-performing settings. 
Overall, \ourtool is robust to moderate configuration changes, and its default design choices are well justified.}
\end{tcolorbox}

\subsection{RQ5: The Costs for Synthesizing Mutators}
\bocomment{In this section, the reported per-mutator cost refers to the end-to-end LLM API cost of the synthesis pipeline, including mutator invention, implementation synthesis, and all refinement attempts, but excluding the one-time fine-tuning cost reported in Section~\ref{subsec:impl}.}
We process a total of 1,000 historical compiler bug reports, including 500 for Rust and 500 for C++, through the full multi-agent pipeline of \ourtool. This process ultimately produces 319 valid mutators for Rust and 403 for C++.

To assess the cost-efficiency of \ourtool, we monitor the number of prompts (LLM inputs) and completions (LLM outputs) generated by each agent throughout the synthesis process. As summarized in \tabref{costs}, the complete pipeline consumes approximately 4.77M and 8.98M LLM input tokens for Rust and C++, respectively, resulting in 1.33M and 1.15M generated responses.

To further understand how individual agents contribute to the overall synthesis efficiency and how each consumes tokens relative to its productive output, we analyze their respective output yields, defined as the ratio of output tokens to input tokens. As shown in \tabref{costs}, the three agents display distinct efficiency characteristics. 
For Rust, the \textit{Mutator Implementation Synthesis} agent accounts for the smallest proportion of token cost (17\%), while for C++, the \textit{Mutator Invention} agent costs the lowest (accounting for 14\%).
The difference is caused by language features and the complexity of mutator implementation.
For both languages, the \textit{Mutator Refinement} agent incurs the highest token costs, accounting for 47\% for Rust and 57\% for C++, respectively.
This is expected as the LLMs have to iteratively correct the invalid mutators, which may involve syntax errors, type mismatches, or even incorrect code logic.
For all agents, the number of input tokens is significantly higher than the number of output tokens, indicating that effective prompt engineering is necessary to reduce overall costs.

We further normalize the cost by the number of valid mutators produced. On average, synthesizing a mutator candidate requires 12,208 LLM tokens for Rust and 20,264 tokens for C++. 
\bocomment{When normalizing by the number of final valid mutators, while treating the cost of invalid and discarded mutators as unavoidable overhead}, \ourtool consumes 19,135 tokens per Rust mutator and 25,142 tokens per C++ mutator.
These figures reflect the total prompt–completion tokens required for automated mutator generation and remain practical in modern development pipelines, especially given the reusability of high-quality mutators across multiple fuzzing campaigns.
Based on GPT-4o pricing in 2025, the estimated \bocomment{end-to-end API cost for producing one final valid mutator is approximately \$0.074 for Rust and \$0.080 for C++, where the estimate includes all generation and refinement attempts amortized over the final valid mutators, but excludes the one-time fine-tuning cost.}
These results indicate that \ourtool provides a cost-effective way to generate a large number of useful mutators efficiently, thereby facilitating early detection of compiler bugs.

\begin{table}[tp]
  \centering
  \small
  \caption{The costs of each agent of \ourtool.}
    \begin{tabular}{|l|l|l|l||l|l|l|}
    \hline
    \textbf{Language} & \multicolumn{3}{c||}{\textbf{Rust}} & \multicolumn{3}{c|}{\textbf{C++}} \\
    \hline
    \textbf{Token Type} & \textbf{Input} & \textbf{Output} & \textbf{Sum} & \textbf{Input} & \textbf{Output} & \textbf{Sum} \\
    \hline
    \hline
    \textbf{Mut. Invent.} & 2,034,827 & 150,476 & 2,185,303 & 1,276,980 & 147,223 & 1,424,203\\
    \hline
    \textbf{Mut. Impl. Synth.} & 749,182 & 290,288 & 1,039,470 & 2,694,638 & 284,096 & 2,978,734 \\
    \hline
    \textbf{Mut. Ref.} & 1,991,245 & 888,007 & 2,879,252 & 5,005,799 & 723,354 & 5,729,153 \\
    \hline
    \hline
    \textbf{Total} & 4,775,254 & 1,328,771 & 6,104,025 & 8,977,417 & 1,154,673 & 10,132,090 \\
    \hline
    \textbf{Total / All Mutators \#} & 9,551 & 2,658 & 12,208 & 17,955 & 2,309 & 20,264 \\
    \hline
    \textbf{Total / Valid Mutators \#} & 14,969 & 4,165 & 19,135 & 22,276 & 2,865 & 25,142 \\
    \hline
    \end{tabular}
  \tablabel{costs}
\end{table}

\begin{tcolorbox}[boxrule=1pt,left=2pt,right=2pt,top=2pt,bottom=2pt]
\underline{\textbf{Answer to RQ5:}} \ourtool incurs low \bocomment{end-to-end synthesis costs.}
On average, producing one final valid mutator requires 14,969/22,276 input tokens and 4,165/2,865 output tokens for Rust/C++, respectively, \bocomment{corresponding to about \$0.074/\$0.080 in API cost per valid mutator, excluding the one-time fine-tuning cost.}
\end{tcolorbox}

\subsection{RQ6: Characteristics and Effectiveness of Synthesized Mutators}
\bocomment{We classify a mutator as valid if it can be successfully compiled, can be applied to the validation test suite without errors, and produces at least one syntactic change on the test suite. Otherwise, it is classified as invalid.}
For Rust, \ourtool generated 64\% valid mutators (319 out of 500). Among the invalid ones, 110 were filtered out for producing no observable mutations on the test suite, and 71 failed to compile after 10 rounds of refinement.
\bocomment{Note that this analysis uses the default refinement setting of 10 rounds. The sensitivity to different refinement-round limits is studied in Section~\ref{sec:eval-refinement-rounds}.}
Notably, \bocomment{among all 500 Rust mutators,} 397 of the Rust mutators required at least one round of refinement, with an average of 3.51 refinement iterations per mutator. 
For C++, \ourtool synthesized 403 valid mutators (81\%), while 9 had no effect against the test suite and 88 failed after 10 rounds of refinement attempts. Among \bocomment{all 500 C++ mutators}, 264 (52.8\%) required at least one refinement attempt, with an average of 2.48 iterations.

We further manually analyze the characteristics of both valid and invalid mutators to better understand their behavior.

\subsubsection{Labeling Process}
To better understand the behavior and quality of synthesized mutators, we conduct a manual inspection over a statistically representative subset of both valid and invalid mutators.
\bocomment{Specifically, we sample mutators at a \textit{95\% confidence level} with a \textit{5\% margin of error}, resulting in 175 Rust and 197 C++ valid mutators, as well as 61 Rust and 72 C++ invalid mutators.}

We apply a structured card sorting methodology to both categories~\cite{yang2024ecosystem,wang2025comprehensive}.
(1) For valid mutators, \bocomment{we analyze both their specification faithfulness and mutation patterns.
First, we check whether each transformation faithfully conforms to the intended mutation specification synthesized in the earlier stage. 
If not, we classify it as a specification mismatch and further categorize the cause. 
Second, we inspect the actual code changes performed by each valid mutator and construct a taxonomy of mutation patterns.}
(2) For invalid mutators, we similarly categorize the reasons why they fail validation. 
Each mutator is independently assessed by at least two annotators, with conflicts resolved through consensus.

\subsubsection{Faithfulness of Valid Mutators}
Two authors manually reviewed each mutator and examined the alignment between code logic and the corresponding specifications.
The analysis showed that 44\% of the Rust mutators and 37\% of the C++ mutators faithfully implemented the intended mutation specifications.
\bocomment{Note that ``faithfully implemented'' is stronger than ``valid'': a mutator may be valid for fuzzing while still deviating from the exact mutation behavior described in its specification.}
Interestingly, among the mutators that triggered confirmed bug discoveries, approximately 65\% were fully aligned with their specifications. 
This suggests that even \textit{partially incorrect mutators can uncover bugs.}
\bocomment{For example, some of these non-faithful mutators are over-constrained or partially implemented variants of the intended transformations, so they may remain applicable on some seed programs and still expose bugs.}
Nonetheless, higher implementation fidelity is generally associated with improved bug-finding ability, underscoring the importance of enhancing mutator correctness in future work.

To further understand the specification mismatches among non-faithful valid mutators, we categorized them into three types. These three categories cover the non-faithful valid mutators in our sampled set.

\begin{enumerate}
    \item \textbf{Incorrect AST Node Matching:} the mutator failed to match the correct AST nodes. For example, the LLM-synthesized mutators usually mistakenly use the variable names that are used in the specification examples, rather than the AST node type of identifier, to locate the target node.
    This type accounts for 11\% and 22\% of the valid mutators for Rust and C++, respectively.
    
    \item  \textbf{Incorrect Mutation:} the mutator incorrectly implements the intended transformation described in the specification. For instance, the specification may require deleting only the right-hand side expression of an assignment statement. However, the LLM-synthesized mutator mistakenly removes all tokens following the assignment operator, including the semicolon at the end of the line, resulting in unintended behavior.
    This type accounts for 31\% and 31\% of the valid mutators for Rust and C++, respectively.
    
    \item \textbf{Missing Steps of Specification:} the specification requires multiple coordinated edits, but the mutator only performs a subset of them.
    For example, a specification may instruct the removal of a parameter from a function signature. However, the LLM‑synthesized mutator may only modify the function signature while failing to remove the parameter’s remaining references, resulting in semantic errors.
    This type accounts for 14\% and 10\% of the valid mutators for Rust and C++, respectively.
\end{enumerate}

\subsubsection{\bocomment{Taxonomy of Mutation Patterns in Valid Mutators}} \label{sec:taxonomy-valid-mutators}

\bocomment{To better understand the design space of the synthesized valid mutators, we manually analyzed the same subset of valid mutators for each language. Each mutator was examined to identify the type of AST modification it performed. Based on these observations, we constructed a taxonomy that categorizes mutators into six classes:
\begin{enumerate}
    \item \textbf{Declaration-level mutations:} modifications to variable, function, or type declarations.
    \item \textbf{Type-structure mutations:} modifications affecting the type hierarchy or structural relationships.
    \item \textbf{Generic / template-related mutations:} modifications involving templates or generic constructs.
    \item \textbf{Constraint / bound-related mutations:} modifications to type constraints, bounds, or template parameters.
    \item \textbf{Expression / attribute-level mutations:} modifications within expressions or attributes.
    \item \textbf{Layout / cast / initialization-related mutations:} modifications affecting object layout, type casting, or initialization.
\end{enumerate}}

\begin{table}[tp]
  \centering
  \small
  \caption{\bocomment{Distribution of valid mutators across mutation categories for Rust and C++.}}
    \begin{tabular}{|l|r|r|}
    \hline
    \bocomment{\textbf{Mutation Category}} & \bocomment{\textbf{Rust}} & \bocomment{\textbf{C++}} \\
    \hline
    \bocomment{Declaration-level mutations} & \bocomment{21} & \bocomment{14} \\
    \bocomment{Type-structure mutations} & \bocomment{32} & \bocomment{29} \\
    \bocomment{Generic / template-related mutations} & \bocomment{49} & \bocomment{73} \\
    \bocomment{Constraint / bound-related mutations} & \bocomment{25} & \bocomment{19} \\
    \bocomment{Expression / attribute-level mutations} & \bocomment{32} & \bocomment{39} \\
    \bocomment{Layout / cast / initialization related mutations} & \bocomment{16} & \bocomment{23} \\
    \hline
    \bocomment{\textbf{Total}} & \bocomment{175} & \bocomment{197} \\
    \hline
    \end{tabular}
  \tablabel{mutator-taxonomy}
\end{table}

\bocomment{\tabref{mutator-taxonomy} summarizes the distribution of valid mutators across these categories for Rust and C++.
For both Rust and C++, a significant portion of mutators (especially generic/template-related and type-structure mutations) target type-level structures, whereas declaration-level and layout-related mutations are less frequent.
This distribution suggests that \ourtool does not merely generate shallow syntactic edits, but can synthesize mutators involving advanced type-level language features.}

\subsubsection{Analysis of Invalid Mutators}
Using the same sampling setups, we analyzed 61 Rust and 72 C++ invalid mutators. 
We identified five primary causes of failure:
\begin{enumerate}
    \item \textit{Use of Deprecated APIs:} the mutator fails because it invokes outdated AST manipulation APIs. These APIs have changed with evolution, but the LLM may have learned from legacy training data. Therefore, it generates transformations using API calls that no longer exist or behave differently, leading to errors.
    This type accounts for 15\% and 67\% of the invalid mutators for Rust and C++, respectively.
    \item \textit{Incorrect Parameters:} the mutator fails due to the misuse of APIs with invalid arguments. 
    This type accounts for 51\% and 17\% of the invalid mutators for Rust and C++, respectively.
    \item \textit{Syntax Errors:} the mutator fails due to syntactic errors within its implementation. 
    This type accounts for 18\% and 8\% of the invalid mutators for Rust and C++, respectively.
    \item \textit{Operand Type Mismatch:} the mutator fails due to the misuse of the operands of incorrect types. 
    This type accounts for 15\% and 1\% of the invalid mutators for Rust and C++, respectively.
    \item \textit{Use of Undeclared Names:} the mutator fails because it uses variables or functions without declarations. This type accounts for 1\% and 7\% of the invalid mutators for Rust and C++, respectively.
\end{enumerate}

The most common failure in Rust was \textit{Incorrect Parameters} (51\%), while in C++ it was the \textit{Use of deprecated APIs} (67\%). These results indicate that synthesizing structurally correct and type-safe AST manipulations remains challenging for advanced LLMs, particularly in languages with verbose or rapidly evolving APIs like Rust.
\bocomment{These results also show that fine-tuning alone cannot fully prevent API-related errors or guarantee faithful transformations, which motivates our validation and refinement stages.}

\begin{tcolorbox}[boxrule=1pt,left=2pt,right=2pt,top=2pt,bottom=2pt]
\underline{\textbf{Answer to RQ6:}} 
44\% of Rust valid mutators and 37\% of C++ valid mutators faithfully implemented their intended specifications.
\bocomment{Our taxonomy further shows that many valid mutators target advanced type-related constructs, especially generic/template-related and type-structure mutations.
Among non-faithful valid mutators, the most common mismatch is \textit{Incorrect Mutation}. Among invalid mutators, the dominant failure causes are \textit{Incorrect Parameters} for Rust and \textit{Use of Deprecated APIs} for C++.}
\end{tcolorbox}

%% file: Discussion.tex
\section{Discussion}
\label{sec:discussion}

In this section, we first discuss the limitations of our study and the implications for future work, then discuss the threats to validity.


\subsection{Limitations and Implications}
Our study demonstrates the effectiveness of leveraging LLMs to invent and implement compiler fuzzing mutators. With carefully designed prompts and systematic validation mechanisms, \ourtool functions as an end-to-end framework, offering a compelling alternative or complement to manually crafted mutators. It can generate a large number of mutators at relatively low cost, significantly reducing human effort and enabling scalable mutation design.

Despite these advantages, our evaluation also reveals several key limitations. First, only about 40\% of the valid mutators fully adhere to their intended specifications. Among those that triggered confirmed bugs, about 65\% were correctly implemented. This suggests that while imperfect mutators may still uncover bugs by chance, higher implementation accuracy generally correlates with better bug-finding effectiveness. Improving the semantic fidelity of generated mutators remains a critical direction for future work.

Second, our analysis of invalid mutators shows that many failures stem from the use of deprecated APIs—likely a result of legacy patterns present in the LLM's training corpus. Future work should explore techniques to proactively detect and avoid such outdated constructs during generation, such as compiler warning filtering, specification-grounded constraints, or fine-tuning on modern codebases.

\bocomment{Finally, although our study demonstrates the cross-language potential of \ourtool, this generality mainly lies at the workflow level rather than in a fully plug-and-play implementation. 
The core pipeline, including \textit{Mutator Invention}, \textit{Mutator Implementation Synthesis}, and \textit{Mutator Refinement}, together with the overall agentic design and prompting strategy, can be reused across different programming languages. 
However, instantiating \ourtool for a new language still requires several language-specific components, including: (1) a language-specific mutator template and corresponding AST transformation APIs, (2) a small set of handwritten mutators for fine-tuning and few-shot guidance, (3) seed programs suitable for the target language, and (4) compiler configurations. 
Therefore, extending \ourtool to new languages does not require redesigning the framework itself, but it does require a moderate setup effort to provide these language-specific bindings and resources.}

\subsection{Threats to Validity}
In this subsection, we discuss the threats to the validity of our study.

\subsubsection{Internal Validity}
\bocomment{The main internal validity threat comes from the stochastic nature of LLM-based generation and the sensitivity of \ourtool to prompt design and configuration choices. Different prompts, temperatures, backend LLMs, fine-tuning examples, or refinement limits may lead to different synthesized mutators and thus affect the final fuzzing results. To mitigate this threat, we use structured prompts with explicit input/output constraints, validate synthesized mutators through compilation and application on the validation test suite, and apply iterative refinement based on compiler feedback. We also conduct extensive ablation and sensitivity studies to examine the effects of key components and configurations, including agent ablation, seed enhancement, temperature, backend LLMs, fine-tuning examples, and refinement rounds. 

Another threat lies in the manual analysis involved in bug confirmation, mutator faithfulness assessment, and mutation-pattern taxonomy construction. Manual judgment may introduce bias, especially when determining whether a mutator faithfully implements its intended specification or whether two crashes correspond to the same underlying bug. To reduce this threat, at least two authors independently inspected the sampled mutators and resolved disagreements through discussion. For mutator analysis, we used statistically representative samples under a 95\% confidence level and 5\% margin of error. For crash comparison, we followed the common practice of comparing the top two stack frames and further manually checked the corresponding programs and compiler outputs.}

\subsubsection{External Validity}
\bocomment{We evaluate \ourtool on Rust and C++, two widely used system-level programming languages, and on four production-level compilers, i.e., rustc, gccrs, GCC, and Clang. 
Although these languages and compilers are representative targets for compiler fuzzing, our results may not directly generalize to other programming languages, especially those with substantially different syntax, type systems, runtime models, or compiler infrastructures. 
Instantiating \ourtool for a new language may require language-specific mutator templates, AST manipulation APIs, seed programs, compiler configurations, and a small set of demonstration mutators.

Another potential threat comes from the selected bug reports and compiler versions. 
We collect historical bug reports from well-maintained open-source repositories and evaluate them on specific compiler versions. 
Different repositories, newer compiler versions, or different distributions of historical bugs may lead to different synthesized mutators and bug-finding results.
To mitigate this threat, we include both Rust and C++ ecosystems, evaluate multiple compilers per language, and further study the impact of the number of analyzed bug reports.
In addition, we used the latest available compiler versions when the experimental campaign started. 
Because production compilers evolve rapidly, newer commits appeared during or after our long-running experiments. 

The choice of backend LLMs may also affect generalizability. 
Our main experiments use GPT-4o, while the performance may vary with other closed-source or open-source models. 
To partially address this threat, we additionally evaluate several alternative backend LLMs and show that the model choice affects the number and effectiveness of synthesized mutators. 
Finally, although our baselines cover both traditional and LLM-based compiler fuzzers, future fuzzers may adopt different designs or stronger models, which could change the comparative results.}

\subsubsection{Construct Validity}
\bocomment{We use several metrics to evaluate \ourtool, including compiler line coverage, unique crashes, confirmed bugs, valid mutator counts, and the number of reproduced confirmed crashes.
These metrics are commonly used or practically meaningful in compiler fuzzing, but they may not fully capture all aspects of fuzzing effectiveness.
For example, line coverage does not necessarily imply semantic diversity or bug discovery, and crash counts do not capture non-crashing miscompilation or logic bugs.
Similarly, the number of valid mutators measures whether synthesized mutators are executable and applicable, but does not guarantee that they faithfully implement the intended specifications.
To mitigate these threats, we complement coverage and crash-based metrics with developer-confirmed bug reports, manual mutator analysis, and a taxonomy of mutation patterns.

Another threat comes from the definition of unique crashes and reproduced bugs.
Following prior work, we identify unique crashes by comparing the top two stack frames, excluding compiler helper functions.
Although this criterion is practical and widely used, it may merge distinct bugs with similar stack traces or separate related bugs with different stack traces.
Therefore, for confirmed-bug reproduction, we further manually inspect the generated programs and compiler outputs.

Finally, fuzzing experiments are subject to randomness and timeouts, which may introduce variability.
We mitigate this threat by running each fuzzer five times and reporting averaged coverage results, while aggregating unique crashes across runs.}

%% file: Related.tex
\section{Related Work}
\label{sec:related}

In this section, we discuss the related work of our study.

\subsection{Compiler Fuzzers}
Compiler fuzzers have uncovered numerous bugs in production-level compilers~\cite{chen2020survey,xu2023silent,marcozzi2019compiler,chen2016empirical}, many of which are critical~\cite{marcozzi2019compiler,zhong2025understanding,he2025bug} and can adversely impact downstream software systems. Broadly, most compiler fuzzing techniques fall into two main categories: \textit{generation-based} and \textit{mutation-based} approaches.

\subsubsection{Generation-based fuzzers}
Generation-based fuzzers synthesize programs from scratch, typically using predefined grammar rules or learned probabilistic models. Tools like CSmith~\cite{yang2011finding} and YARPGen~\cite{livinskii2020random,livinskii2023fuzzing} exemplify this category, generating semantically valid programs to explore diverse compiler behaviors.
\bocomment{In addition, some semantics-aware fuzzers encode typing rules or logic constraints to generate well-typed programs~\cite{dewey2015fuzzing}.}
More recent work has extended this paradigm into other languages, including Java~\cite{chen2023compiler,zhao2022history}, JavaScript~\cite{ye2023generative}, Rust~\cite{sharma2023rustsmith,wang2024rustlantis},  MLIR~\cite{wang2023mlirsmith}, deep-learning compilers~\cite{wang2023gencog,liu2023nnsmith,ren2025effective}, SQL executors~\cite{hua2023gdsmith}, and Kotlin~\cite{georgescu2024evolutionary,chaliasos2022finding}.
Researchers also proposed auxiliary approaches to boost generation-based fuzzers~\cite{even2020closer,even2022csmithedge,wu2023boosting,li2024boosting}.

\bocomment{More recently, LLM-based generation fuzzers have been explored for compiler testing~\cite{yang2024whitefox,gao2025clozemaster,ni2025interleaving,xia2024fuzz4all}. For example, Clozemaster~\cite{gao2025clozemaster} fine-tunes an LLM to complete masked code segments, while Fuzz4All~\cite{xia2024fuzz4all} uses LLMs to rewrite seed programs guided by language standards and API references. 
These methods use LLMs to generate or edit individual test programs, whereas \ourtool uses LLMs to synthesize reusable mutation operators.}


\bocomment{While generation-based fuzzers can produce valid programs when sufficient semantic knowledge is encoded, supporting complex language features such as generics, trait bounds, ownership, templates, or compiler-specific extensions often requires substantial manual modeling effort. 
In contrast, \ourtool follows a mutation-based design: it reuses real-world seed programs and compiler test cases, thereby preserving much of their existing syntactic and semantic context while exploring new compiler behaviors through synthesized mutators.}

\subsubsection{Mutation-based fuzzers}
Mutation-based fuzzers generate program variants by applying transformations to seed programs. Representative approaches include EMI-based fuzzers~\cite{le2014compiler,le2015finding,gao2024shoot,theodoridis2022finding,sun2016hermes} and differential fuzzers targeting C/C++~\cite{even2023grayc,jiang2021ctos,liu2023flux,wang2025fuzzing,wu2025unveiling,wu2026stochastic}, JIT languages~\cite{chen2016coverage,chen2019deep,schumi2021spectest,wu2023jitfuzz,park2023feature,zang2023jattack,xie2024validating,feng2025finding}, deep learning compilers~\cite{li2023validating,chen2025scuzer}, MLIR~\cite{suo2024fuzzing}, Rust~\cite{yang2024rust}, \bocomment{Swift~\cite{hyatt2026mutation}}, and various domain-specific languages~\cite{tu2022detecting,bendrissou2024grammar}.
These methods typically rely on manually designed mutators, requiring significant domain expertise and engineering effort.

MetaMut~\cite{ou2024mutators} represents the first attempt to synthesize mutators using LLMs, composing transformations from predefined AST-level operations. 
However, its search space is constrained by these predefined operations, and over half of the generated mutators require manual repair to function correctly.
\bocomment{Different from MetaMut, \ourtool derives mutator specifications from historical compiler bug reports and then synthesizes feature-specific mutator implementations through a multi-agent pipeline consisting of mutator invention, implementation synthesis, and refinement. 
This design enables \ourtool to generate mutators targeting language features and failure patterns observed in real compiler bugs, rather than relying only on direct program rewriting or fixed transformation primitives.}


\subsection{LLM-Based Testing}
Recent advances have demonstrated the potential of LLMs in automated software testing~\cite{wang2024software,chen2025deep}.
Many works focus on generating unit tests, including test functions~\cite{schafer2023empirical,chen2024chatunitest,yang2024evaluation,yuan2024evaluating,shang2025large}, input values~\cite{chen2024reasoning}, and assertion oracles~\cite{hossain2025doc2oracll,molina2025test}, often guided by code structure, comments, or coverage feedback.
\bocomment{Bug reports are used to enhance test generation~\cite{ouedraogo2025enriching,shiri2025history} or configure generation~\cite{huangoptfuzz}}.
These approaches primarily operate at the unit-test level, aiming to validate method-level behavior, and their outputs are usually test code written in the same language as the program under test.
In contrast, our work targets compiler testing, focusing on mutation operators that systematically mutate code to explore deep compiler behavior.

%% file: Conclusion.tex
\section{Conclusion}
\label{sec:conclusion}

We present \ourtool, a cross-language framework \bocomment{for automated mutation-operator synthesis} using LLMs guided by historical bug reports.
\ourtool involves three LLM agents, namely the \textit{Mutator Invention} agent, the \textit{Mutator Implementation Synthesis} agent, and the \textit{Mutator Refinement} agent.
\bocomment{In our main evaluation with 500 bug reports per language,} \ourtool generates 319 and 403 valid mutators for Rust and C++, respectively.
\bocomment{We further study the impact of the bug-report budget by analyzing 200 additional reports per language, which contribute 125 and 158 additional valid mutators for Rust and C++, respectively.}
Moreover, we expand the input seed programs with our adaptive seed enhancement approach.

In our bug-hunting experiment \bocomment{using 700 bug reports per language}, \ourtool reported \bocomment{100} compiler bugs across four production compilers, \bocomment{61} of which were confirmed as previously unknown and \bocomment{35} of which have already been fixed.
\bocomment{Our ablation and sensitivity studies further demonstrate the necessity of the key agents, the usefulness of seed enhancement, and the robustness of the default configurations. The low end-to-end API cost of synthesizing valid mutators demonstrates the promise of LLM-enhanced pipelines as a practical alternative to manual mutator development.}

In the future, we plan to reduce invalid mutators and enhance the correctness of mutator implementation by symbolic reasoning or reinforcement learning.
\textbf{Our artifact is publicly available at \url{https://github.com/sososopy/Mut4All-Artifacts}.}